\documentclass[preprintnumbers,prd,showpacs,floatfix,superscriptaddress,nofootinbib,twocolumn, letterpaper]{revtex4-1}
\usepackage{longtable}
\usepackage{bm}
\usepackage{relsize}
\usepackage{amsfonts}
\usepackage{amsmath}
\usepackage{amssymb,epsf}
\usepackage{latexsym}
\usepackage{graphicx,epsfig}
\usepackage{amssymb}
\usepackage{float}
\usepackage{subfigure}
\usepackage{epstopdf}
\usepackage[colorlinks=true,citecolor=blue,linkcolor=blue,urlcolor=black]{hyperref}
\usepackage{dcolumn}
\usepackage{psfrag}
\usepackage{wrapfig}
\usepackage{makeidx}
\usepackage{epsf}
\usepackage{color}
\usepackage{multirow}
\usepackage{mathtools}
\begin{document}
%%%%%%%%%%%%%%%%%%%%%%%%%%%%%%%%%%%%%%%%%%%%%%%%%%%%%%%%%%%%%%%%%%%%%%%%%%%%

\title{Cosmological sudden singularities in $f(R,T)$ gravity}

\author{Tiago B. Gon\c{c}alves}
\email{tgoncalves@alunos.fc.ul.pt}
\affiliation{Instituto de Astrof\'{i}sica e Ci\^{e}ncias do Espa\c{c}o, Faculdade de Ci\^{e}ncias da Universidade de Lisboa, Edif\'{i}cio C8, Campo Grande, P-1749-016 Lisbon, Portugal}
\affiliation{Departamento de F\'{i}sica, Faculdade de Ci\^{e}ncias da Universidade de Lisboa, Edif\'{i}cio C8, Campo Grande, P-1749-016 Lisbon, Portugal}

\author{Jo\~{a}o Lu\'{i}s Rosa}
\email{joaoluis92@gmail.com}
\affiliation{Institute of Physics, University of Tartu, W. Ostwaldi 1, 50411 Tartu, Estonia}

\author{Francisco S. N. Lobo}
\email{fslobo@fc.ul.pt}
\affiliation{Instituto de Astrof\'{i}sica e Ci\^{e}ncias do Espa\c{c}o, Faculdade de Ci\^{e}ncias da Universidade de Lisboa, Edif\'{i}cio C8, Campo Grande, P-1749-016 Lisbon, Portugal}
\affiliation{Departamento de F\'{i}sica, Faculdade de Ci\^{e}ncias da Universidade de Lisboa, Edif\'{i}cio C8, Campo Grande, P-1749-016 Lisbon, Portugal}
\date{\today}

%%%%%%%%%%%%%%%%%%%%%%%%%%%%%%%%%%%%%%%%%%%%%%%%%%%%%%%%%%%%%%%%%%%%%%%%
\begin{abstract} 
In this work, we study the possibility of finite-time future cosmological singularities appearing in $f(R,T)$ gravity, where $R$ is the Ricci scalar and $T$ is the trace of the stress-energy tensor. We present the theory in both the geometrical and the dynamically equivalent scalar-tensor representation and obtain the respective equations of motion. In a background Friedmann-Lema\^{i}tre-Robertson-Walker (FLRW) universe with an arbitrary curvature and for a generic $C^\infty$ function $f(R,T)$, we prove that the conservation of the stress-energy tensor prevents the appearance of sudden singularities in the cosmological context at any order in the time-derivatives of the scale factor. However, if this assumption is dropped, the theory allows for sudden singularities to appear at the level of the third time-derivative of the scale factor $a(t)$, which are compensated by divergences in either the first time-derivatives of the energy density $\rho(t)$ or the isotropic pressure $p(t)$. For these cases, we introduce a cosmological model featuring a sudden singularity that is consistent with the current measurements for the cosmological parameters, namely, the Hubble constant, deceleration parameter, and age of the universe, and provide predictions for the still unmeasured jerk and snap parameters. Finally, we analyse the constraints on a particular model of the function $f(R,T)$ that guarantees that the system evolves in a direction favorable to the energy conditions at the divergence time.
\end{abstract}
%%%%%%%%%%%%%%%%%%%%%%%%%%%%%%%%%%%%%%%%%%%%%%%%%%%%%%%%%%%%%%%%%%%%%%%%

\pacs{04.50.Kd, 04.20.Cv}

\maketitle

%\tableofcontents

%%%%%%%%%%%%%%%%%%%%%%%%%%%%%%%%%%%%%%%%%%%%%%%%%%%%%%%%%%%%%%%%%%%%%%%%
\section{Introduction}\label{sec:intro}
%%%%%%%%%%%%%%%%%%%%%%%%%%%%%%%%%%%%%%%%%%%%%%%%%%%%%%%%%%%%%%%%%%%%%%%%

What will be the fate of our universe? Our usual answer depends on the curvature parameter: if the universe is hyperspherical (with curvature parameter $k=1$) then it may reach a maximum of expansion and recollapse ending in a `Big Crunch'; or, if it is spatially flat ($k=0$) or hyperbolic ($k=-1$) then it will expand forever in a `Big Freeze', becoming cooler and emptier (while `islands' of galaxies may still be able to survive). However, as with any dynamical physical system, such predictions are naive, as we do not have a precise knowledge of the present conditions such as of the spatial inhomogeneities and of the composition of the universe; or there could be transformations between different kinds of matter in the future, which could all play a decisive role in the evolution of the universe \cite{Starobinsky:1999yw}. 

Surprisingly, we now understand the expansion of the universe to be accelerating \cite{SupernovaCosmologyProject:1998vns,SupernovaSearchTeam:1998fmf}. A possible explanation, which fits the data well, is that this acceleration is driven by an exotic fluid with an equation of state $w=-1$, consistent with a cosmological constant. But could it be evolving, for instance, into the phantom regime ($w<-1$)? If so, the end could be much more dramatic with a singularity (e.g. a divergence of the scale factor) occurring at a finite, future time. In this case, the universe would end in a `Big Rip', with galaxies, planets and atoms being ripped apart by an ever increasing domination of this dark (phantom) energy \cite{Caldwell:2003vq,Gonzalez-Diaz:2003xmx,Gonzalez-Diaz:2003bwh,Bouhmadi-Lopez:2006fwq}. Finite-time, future singularities have been classified in four types depending on which parameters diverge at the time of the singularity $t_s$ \cite{Nojiri:2005sx}: 
\begin{itemize}
	\item Type I (`Big Rip'): For $t\rightarrow t_s$, $a\rightarrow \infty$, $\rho\rightarrow\infty$ and $\left|p\right|\rightarrow\infty$. That is, as we approach the finite time of the singularity $t_s$, the scale factor $a$ diverges. Even if the energy density $\rho$ and pressure $p$ do not diverge at this time, the singularity will still be included in this type. 
	
	\item Type II (`Sudden'): For $t\rightarrow t_s$, $a\rightarrow a_s$, $\rho\rightarrow \rho_s$ and $\left|p\right|\rightarrow\infty$. In this case, at $t_s$ both $a$ and $\rho$ remain finite ($a_s\equiv a\left(t_s\right)$ and $\rho_s\equiv \rho\left(t_s\right)$ are finite constants), and it is the pressure $p$ which diverges. Usually it is considered that the expansion rate $H\equiv \dot{a}/a$ remains finite \cite{Barrow:2004hk}, but there may be divergences in higher derivatives of $a$. 
	
	\item Type III: For $t\rightarrow t_s$, $a\rightarrow a_s$, $\rho\rightarrow\infty$ and $\left|p\right|\rightarrow\infty$. That is, in this case, the energy density $\rho$ also diverges.
	
	\item Type IV: For $t\rightarrow t_s$, $a\rightarrow a_s$, $\rho\rightarrow 0$ and $p\rightarrow 0$. Here, the singularity appears at $t_s$ due to divergences in higher derivatives of $a$ (or higher derivatives of the Hubble expansion rate $H\equiv \dot{a}/a$). It also includes the cases in which $\rho$ and $p$ tend to some finite value. 
\end{itemize}
For other types of singularities, such as `Big Brake' and `Big Separation', see also Refs.~\cite{Chimento:2015gum,Chimento:2015gga,Cataldo:2017nck}.

What else could be the cause for the accelerated expansion of the universe? An alternative explanation is to consider that general relativity (GR) breaks down at cosmological scales, in which case we need a modified theory of gravity \cite{Nojiri:2006ri,Lobo:2008sg,Nojiri:2010wj,Clifton:2011jh,Capozziello:2011et,CANTATA:2021ktz,Avelino:2016lpj} to describe the dynamics at these large scales. For instance, one can consider the gravitational Lagrangian to be dependent on a general function of the Ricci curvature scalar $R$, instead of depending linearly on $R$ as in the Einstein-Hilbert action of GR. This simple modification is the so-called $f(R)$ gravity \cite{Sotiriou:2008rp}, which has been shown to be consistent with an accelerated expansion without necessarily requiring a dark energy component \cite{Capozziello:2002rd}. Different approaches may be taken within a gravity theory. For instance, various formalisms have been used in $f(R)$ gravity: the metric formalism which consists in varying the action with respect to the metric \cite{Sotiriou:2008rp}, the metric-affine formalism where the metric and the connections are treated as separate variables \cite{Olmo:2011uz} and the hybrid formalism \cite{Harko:2011nh,Harko:2020ibn,Harko:2018ayt,Capozziello:2012ny,Capozziello:2013uya,Capozziello:2015lza,Rosa:2017jld,Rosa:2019ejh,Rosa:2021ish} which unifies the above-mentioned approaches.

The $f(R)$ modification of gravity can be further extended, and one can consider other couplings to matter (beyond the minimal coupling in GR) \cite{Bertolami:2007gv,Harko:2011kv,Harko:2012hm,Harko:2010mv,Harko:2014gwa,Haghani:2013oma,Odintsov:2013iba}. For instance, allowing the gravitational Lagrangian to depend on a general function not only of $R$ but also of the trace of the stress-energy tensor $T$ is a modification known as $f(R,T)$ gravity \cite{Harko:2011kv}. The $T$-dependence in the Lagrangian may arise, for instance, in models of interacting dark energy where a cosmological term in the gravitational Lagrangian is a function of the trace of the stress-energy, $\Lambda(T)$ \cite{Poplawski:2006ey}. Interestingly, because of the explicit coupling between geometry and matter, the matter stress-energy tensor need not be conserved in general (i.e., it allows ${\nabla_\nu T^{\mu\nu} \neq 0}$), thus it could lead to nongeodesic motion of particles and to matter creation from gravitational fields \cite{Harko:2014pqa,Harko:2015pma,Harko:2021bdi}, which could have a semiclassical effective interpretation of quantum effects \cite{Harko:2021tav}. The astrophysical and cosmological applications of $f(R,T)$ gravity have received extensive attention in the literature (see Ref. \cite{Harko:2018ayt} for more details). Recently, an equivalent dynamical scalar-tensor representation of $f(R,T)$ gravity was introduced in Ref.~\cite{Rosa:2021teg} to study junction conditions for the matching between two spacetimes at a separation hypersurface. Thick brane solutions \cite{Rosa:2021tei,Rosa:2021myu,Rosa:2022fhl}, and reconstructed background cosmological solutions \cite{Goncalves:2021vci} have also been explored in the scalar-tensor representation of $f(R,T)$ gravity.

One can also pose the question: How different may the fate of the universe be if it is ruled by modified gravity? Indeed, in the context of the future singularities mentioned above, the future evolution of the universe has been extensively analyzed in $f(R)$ gravity \cite{Nojiri:2005sx,Odintsov:2015zza}. Furthermore, type II, sudden singularities have also been studied in an expanding Friedmann universe \cite{Barrow:2004xh}, in Brans-Dicke theory \cite{Barrow:2019cuv} and in generalized hybrid metric-Palatini gravity \cite{Rosa:2021ish}. In fact, much work has been explored in the literature relative to the finite time, future singularities, for instance, in $f({\cal T})$ gravity, where ${\cal T}$ is the torsion scalar \cite{Bamba:2012vg}, and in string-inspired scalar-Gauss-Bonnet and modified Gauss-Bonnet theories \cite{Bamba:2008ut}, where finite-time future singularities  were found. The latter analysis was extended to $f(R,G)$ gravity \cite{Bamba:2010wfw}, where $G$ is the Gauss-Bonnet invariant, and it was shown that in accelerating cosmologies finite-time future singularities do indeed emerge. However, it was shown explicitly that taking into account the back-reaction of conformal quantum fields near the singularity, quantum effects may delay, or render milder, the singularity \cite{Nojiri:2004ip}. In fact, it was argued that if the evolution to the singularity is realistic, due to quantum effects the universe may end up in a de-Sitter phase before the scale factor diverges.

The main aim of this work is to explore finite-time future singularities in $f(R,T)$ gravity, in both of its geometrical and scalar-tensor representations. Section~\ref{sec:fRTintro} reviews the essentials of both representations of $f(R,T)$ gravity. In Section~\ref{sec:cosmo}, it is assumed that the stress-energy tensor is conserved: \ref{subsec:framework} explains the assumptions and the framework under which our work is done;  \ref{subsec:typeII-geometrical} and \ref{subsec:typeII-scalar-tensor} show the absence of type II, sudden singularities, both in the geometrical and the scalar-tensor representations, respectively; \ref{subsec:typeIV} considers the absence of type IV singularities. Section~\ref{sec:suddens}, on de other hand, drops the assumption of the stress-energy conservation: subsections~\ref{subsec:sudden-ddda-geom} and \ref{subsec:sudden-ddda-ST}, respectively in the geometrical and the scalar-tensor representations, show that sudden singularities can arise in the third time-derivative of the scale factor; \ref{subsec:modelsingularity} studies a model for the scale factor where such singularities may appear; \ref{subsec:cosmo-constraints} imposes constraints on that model from the measured Hubble constant, deceleration parameter and age of the universe; and \ref{subsec:E-conditions} constrains a particular ${f(R,T)}$ model from considerations on the energy conditions. A summary of our findings can be found in Section~\ref{sec:conclusion}.

%%%%%%%%%%%%%%%%%%%%%%%%%%%%%%%%%%%%%%%%%%%%%%%%%%%%%%%%%%%%%%%%%%%%%%%%
\section{Theory and equations of the $f\left(R,T\right)$ gravity}\label{sec:fRTintro}
%%%%%%%%%%%%%%%%%%%%%%%%%%%%%%%%%%%%%%%%%%%%%%%%%%%%%%%%%%%%%%%%%%%%%%%%
\subsection{Geometrical representation}
%%%%%%%%%%%%%%%%%%%%%%%%%%%%%%%%%%%%%%%%%%%%%%%%%%%%%%%%%%%%%%%%%%%%%%%%

The action $S$ that describes the $f\left(R,T\right)$ gravity theory \cite{Harko:2011kv} is of the form
\begin{equation}\label{eq:fRTaction-original}
    S = \frac{1}{2\kappa^2} \int_{\Omega}\sqrt{-g} f(R,T) d^4 x+ \int_{\Omega} \sqrt{-g} \mathcal{L}_m d^4 x,
\end{equation}
where $\kappa^2=8\pi G/c^4$, where $G$ is the gravitational constant and $c$ is the speed of light, $\Omega$ is the 4-dimensional spacetime manifold on which the set of coordinates $x^\mu$ is defined, $g$ is the determinant of the metric $g_{\mu\nu}$ with a positive signature, $f\left(R,T\right)$ is an arbitrary function of the Ricci scalar $R=g^{\mu\nu}R_{\mu\nu}$, with $R_{\mu\nu}$ the Ricci tensor, and the trace of the stress-energy tensor $T=g^{\mu\nu}T_{\mu\nu}$. The latter is defined in terms of the variation of the matter Lagrangian $\mathcal L_m$ with respect to the metric as
\begin{equation}
T_{\mu\nu}=-\frac{2}{\sqrt{-g}}\frac{\delta\left(\sqrt{-g}\mathcal L_m\right)}{\delta g^{\mu\nu}}.
\end{equation}
From this point onwards, we adopt a geometrized unit system in which $G=c=1$, thus implying $\kappa^2=8\pi$. 

The modified field equations of the $f\left(R,T\right)$ gravity theory can be obtained by taking a variation of Eq.~\eqref{eq:fRTaction-original} with respect to the metric $g_{\mu\nu}$, yielding
\begin{equation}\label{eq:Gfields}
\begin{multlined}
    f_R R_{\mu\nu}-\frac{1}{2}g_{\mu\nu}f(R,T) + \left(g_{\mu\nu}\square-\nabla_\nu\right)f_R \\ 
    = \kappa^2 T_{\mu\nu}-f_T (T_{\mu\nu}+\Theta_{\mu\nu}),
\end{multlined}
\end{equation}
where we have defined $f_R\equiv\partial f/\partial R$ and $f_T\equiv\partial f/\partial T$, $\nabla_\mu$ is the covariant derivative defined in terms of the metric $g_{\mu\nu}$,  $\square\equiv\nabla^\sigma\nabla_\sigma$ is the d’Alembert operator, and $\Theta_{\mu\nu}$ is a tensor defined as
\begin{equation}\label{eq:Theta-varT}
    \Theta_{\mu\nu}\equiv g^{\rho\sigma}\frac{\delta T_{\rho\sigma}}{\delta g^{\mu\nu}}.
\end{equation}
One can take the trace of the field equations in Eq.~\eqref{eq:Gfields} to obtain a relation between $R$ and $T$ of the form:
\begin{equation}\label{eq:traceFieldEqs-Geo}
f_R R-2f+3\square f_R = \left(\kappa^2- f_T\right) T - f_T \Theta,
\end{equation}
where $\Theta= g^{\mu\nu}\Theta_{\mu\nu}$ is the trace of $\Theta_{\mu\nu}$.

Finally, we obtain the conservation equation for $f\left(R,T\right)$ gravity, taking into account the divergence of Eq.~\eqref{eq:Gfields} and using the identity $\left(\square\nabla_\nu-\nabla_\nu\square\right)f_R=R_{\mu\nu}\nabla^\mu f_R$. The result is as follows
\begin{equation}\label{eq:conserv-general}
\begin{multlined}
    (\kappa^2-f_T)\nabla^\mu T_{\mu\nu}=\left(T_{\mu\nu}+\Theta_{\mu\nu}\right)\nabla^\mu f_T \\ 
    +f_T\nabla^\mu\Theta_{\mu\nu}+f_R \nabla^\mu R_{\mu\nu}-\frac{1}{2}g_{\mu\nu}\nabla^\mu f.
\end{multlined}
\end{equation}

%%%%%%%%%%%%%%%%%%%%%%%%%%%%%%%%%%%%%%%%%%%%%%%%%%%%%%%%%%%%%%%%%%%%%%%%
\subsection{Scalar-tensor representation}\label{subsec:scalar-tensor}
%%%%%%%%%%%%%%%%%%%%%%%%%%%%%%%%%%%%%%%%%%%%%%%%%%%%%%%%%%%%%%%%%%%%%%%%

Similarly to what happens in other modified theories of gravity where extra scalar degrees of freedom in comparison to GR are featured, it is possible and frequently useful to consider a dynamically equivalent scalar-tensor representation of the $f\left(R,T\right)$ gravity theory, in this case with two scalar fields \cite{Rosa:2021teg}.

In this scalar-tensor representation, the arbitrary dependence of $f\left(R,T\right)$ in the scalars $R$ and $T$ is exchanged by two scalar fields $\varphi$ and $\psi$ and an arbitrary interaction potential $V\left(\varphi,\psi\right)$. These quantities are defined in terms of $f\left(R,T\right)$ and its partial derivatives as
 \begin{equation}\label{eq:varphi&psi}
     \varphi\equiv\frac{\partial f}{\partial R} ,\qquad
    \psi\equiv\frac{\partial f}{\partial T},
 \end{equation}
\begin{equation}\label{eq:potential}
    V(\varphi,\psi) \equiv -f(R,T)+ \varphi R + \psi T.
\end{equation}
Inserting these definitions into Eq. \eqref{eq:fRTaction-original}, one obtains the action of the equivalent scalar-tensor representation of $f\left(R,T\right)$ in the form
\begin{equation}\label{eq:STaction}
    \begin{split} 
    S = \frac{1}{2\kappa^2} \int_{\Omega} \sqrt{-g} \left[\varphi R+\psi T - V(\varphi, \psi)\right]d^4 x \\ 
    + \int_{\Omega} \sqrt{-g} \mathcal{L}_m d^4 x .
    \end{split}
\end{equation}
However, this scalar-tensor representation is only well posed if ${f_{RR}f_{TT}\neq f_{RT}^2}$ \cite{Rosa:2021teg}. Similarly to what happens in the metric approach to $f(R)$ theories of gravity, the scalar field $\varphi$ is analogous to a Brans-Dicke scalar field with parameter $\omega_{BD}=0$ and with an interaction potential $V$. The scalar degree of freedom associated with the dependence in $T$ is carried by the scalar field $\psi$.  

The new action in Eq.~\eqref{eq:STaction} depends now on three independent quantities, namely the metric $g_{\mu\nu}$ and the two scalar fields $\varphi$ and $\psi$. The corresponding modified field equations are again obtained via a variation with respect to $g_{\mu\nu}$, and is given by
\begin{equation}\label{eq:fields}
    \begin{multlined}
      \varphi R_{\mu\nu}-\frac{1}{2}g_{\mu\nu}\left(\varphi R + \psi T - V\right)\\+(g_{\mu\nu}\square-\nabla_\mu\nabla_\nu)\varphi = \kappa^2 T_{\mu\nu} -\psi (T_{\mu\nu} + \Theta_{\mu\nu}),
      \end{multlined}
\end{equation}
which could also be obtained directly from Eq.~\eqref{eq:Gfields} via the introduction of the definitions in Eqs.~\eqref{eq:varphi&psi} and \eqref{eq:potential}. The trace of the field equations in this representation yields the following relation between $R$ and $T$:
\begin{equation}\label{eq:traceFieldEqs-ST}
3\square\varphi-\varphi R + 2 V = (\kappa^2+\psi)T-\psi\Theta.
\end{equation}
Furthermore, the equations of motion for the field $\varphi$ and $\psi$ can be obtained by taking the variation of the action in Eq.~\eqref{eq:STaction} with respect to these scalar fields, respectively, from which one obtains
\begin{equation}\label{eq:Vphi}
    V_{\varphi} = R,
\end{equation}
\begin{equation}\label{eq:Vpsi}
    V_{\psi} = T,
\end{equation}
where we have defined $V_\varphi\equiv \partial V/\partial\varphi$ and $V_\psi\equiv \partial V/\partial\psi$.

Finally, the conservation equation in this representation can be obtained by introducing the definitions of Eqs.~\eqref{eq:varphi&psi} and \eqref{eq:potential} into Eq.~\eqref{eq:conserv-general}, and recalling that the Einstein tensor is divergenceless (Bianchi identities), i.e., $\nabla^\mu\left(R_{\mu\nu}-\frac{1}{2}g_{\mu\nu}R\right)=0$ which leads to
\begin{equation}\label{eq:conserv-general2}
\begin{multlined}
    (\kappa^2-\psi)\nabla^\mu T_{\mu\nu}=\left(T_{\mu\nu}+\Theta_{\mu\nu}\right)\nabla^\mu\psi+ \\ +\psi\nabla^\mu\Theta_{\mu\nu}-\frac{1}{2}g_{\mu\nu}\left[R\nabla^\mu\varphi+\nabla^\mu\left(\psi T-V\right)\right].
\end{multlined}
\end{equation}

%%%%%%%%%%%%%%%%%%%%%%%%%%%%%%%%%%%%%%%%%%%%%%%%%%%%%%%%%%%%%%%%%%%%%%%%
\section{Absence of sudden singularities with $\nabla_\nu T^{\mu\nu}=0$}\label{sec:cosmo}
%%%%%%%%%%%%%%%%%%%%%%%%%%%%%%%%%%%%%%%%%%%%%%%%%%%%%%%%%%%%%%%%%%%%%%%%
\subsection{Framework and assumptions}\label{subsec:framework}
%%%%%%%%%%%%%%%%%%%%%%%%%%%%%%%%%%%%%%%%%%%%%%%%%%%%%%%%%%%%%%%%%%%%%%%%

In this work, we assume that the universe is well-described by an homogeneous and isotropic Friedmann-Lema\^{i}tre-Robsertson-Walker (FLRW) spacetime, which in the usual spherical coordinates $(t,r,\theta,\phi)$ takes the form
\begin{equation}\label{eq:FLRW-metric}
    ds^2 = -dt^2+a^2(t)\left[\frac{dr^2}{1-kr^2}+r^2\left(d\theta^2+\sin^2\theta d\phi^2\right)\right], 
\end{equation}
where $a(t)$ is the scale factor and $k$ is the curvature parameter which can take the values $k=\left\{-1,0,1\right\}$ corresponding to a hyperbolic, spatially flat, or hyperspherical universe, respectively. Furthermore, we also assume that matter is well-described by an isotropic perfect fluid, i.e., the stress-energy tensor $T_{\mu\nu}$ can be written in the form
\begin{equation}\label{eq:em-fluid}
    T_{\mu\nu}=(\rho+p)u_\mu u_\nu +p g_{\mu\nu},
\end{equation}
where $\rho$ is the energy density, $p$ is the isotropic pressure, and $u^\mu$ is the fluid 4-velocity satisfying the normalization condition $u_\mu u^\mu=-1$. Taking the matter Lagrangian to be $\mathcal L_m=p$ \cite{Bertolami:2008ab}, the tensor $\Theta_{\mu\nu}$ takes the form
\begin{equation}\label{eq:Theta-fluid}
    \Theta_{\mu\nu}=-2T_{\mu\nu}+p g_{\mu\nu}.
\end{equation}
To preserve the homogeneity and isotropy of the solutions, all physical quantities introduced are assumed to depend solely on the time coordinate $t$, i.e., $\rho=\rho\left(t\right)$, $p=p\left(t\right)$, $\varphi=\varphi\left(t\right)$, and $\psi=\psi\left(t\right)$. 

Even though the conservation of the stress-energy tensor $T_{\mu\nu}$ is not a mandatory condition in $f\left(R,T\right)$ gravity, allowing thus for transformations of energy between the matter and the extra degrees of freedom of the gravitational sector, we will begin our analysis considering the case where matter is conserved, i.e., we assume that ${\nabla_\nu T^{\mu\nu}=0}$, and we will prove that no sudden singularities arise in this case. In a following section, Sec. \ref{sec:suddens} we will extend the analysis for the general case. Thus, from ${\nabla_\nu T^{\mu\nu}=0}$, one obtains the usual matter conservation equation
\begin{equation}\label{eq:conserv-m}
\dot{\rho} = -3\frac{\dot{a}}{a}(p+\rho),
\end{equation}
where overdots ($\dot{\ }$) denote derivatives with respect to the time coordinate $t$.

It is assumed that the function $f(R,T)$ is a smooth, infinitely differentiable function, i.e., a $C^{\infty}$ function, that admits a Taylor series expansion. Thus, its partial derivatives $f_R=\varphi$ and $f_T=\psi$, and higher order partial derivatives, always remain finite. This assumption has important implications on the quantities $R$ and $T$ themselves. Consider an arbitrary function ${f\left(x\right)}$ that admits a Taylor-series expansion, and a time-varying quantity $x\left(t\right)$. With these two quantities, one can always introduce a composite function ${g\left(t\right) = f\left(x\left(t\right)\right)}$. Now, if $x$ diverges in a finite time $t_s$ and the function $f(x)$ is unbounded, the function $g(t)$ will consequently diverge in the same instant $t_s$. There are two possible ways to prevent the divergence in $g\left(t\right)$: either the function $f\left(x\right)$ is bounded, or the quantity $x\left(t\right)$ remains finite. Since the first of these assumptions would break the arbitrariness of the function $f\left(x\right)$, one must require $x\left(t\right)$ to remain finite throughout its time evolution. This argument can be extrapolated for a function of two variables, e.g., $f\left(R(t),T(t)\right)$, from which one concludes that the smoothness and finiteness of the arbitrary function $f\left(R,T\right)$ implies that $R(t)$ and $T(t)$ must remain finite for all $t$. As the potential $V(\varphi,\psi)$ can be constructed solely form the function $f\left(R,T\right)$ and its partial derivatives, this quantity is also a $C^{\infty}$ function and all its partial derivatives also remain finite.

In the following sections we study whether sudden singularities can appear in either the geometrical or the scalar-tensor representations of the $f(R,T)$ gravity. A sudden singularity could occur if the scale factor $a$, the expansion rate $H=\dot{a}/a$ and the energy density $\rho$ remain finite throughout the entire time evolution, but the pressure $p$ and/or higher derivatives of the scale factor diverge at some finite future time instant $t_s$ \cite{Barrow:2004xh,Barrow:2004hk}, where the subscript $s$ will be used in what follows to denote the values of the quantities at the sudden singularity.

%%%%%%%%%%%%%%%%%%%%%%%%%%%%%%%%%%%%%%%%%%%%%%%%%%%%%%%%%%%%%%%%%%%%%%%%
\subsection{Geometrical representation}\label{subsec:typeII-geometrical}
%%%%%%%%%%%%%%%%%%%%%%%%%%%%%%%%%%%%%%%%%%%%%%%%%%%%%%%%%%%%%%%%%%%%%%%%

Under the assumptions detailed above, Eq.~\eqref{eq:Gfields} features two independent components corresponding to the modified Friedmann equation and the modified Raychaudhuri equation, which take the forms
\begin{equation}\label{eq:Gtt}
	-3f_R\frac{\ddot{a}}{a}+\frac{1}{2}f+3\dot{f_R}\frac{\dot{a}}{a}=8\pi\rho+f_T \left(p+\rho\right),
\end{equation}
\begin{equation}\label{eq:Grr}
	f_R\left(\frac{\ddot{a}}{a}+2\frac{\dot{a}^2+k}{a^2}\right)-\frac{1}{2}f-2\dot{f_R}\frac{\dot{a}}{a}-\ddot{f_R}=8\pi p,
\end{equation}
respectively. Furthermore, the Ricci scalar and the trace of the stress-energy tensor are given by
\begin{equation}\label{eq:R}
R= 6\left( \frac{\ddot{a}}{a}+ \frac{\dot{a}^{2}+k}{a^{2}}\right),
\end{equation}
\begin{equation}\label{eq:T}
T= -\rho + 3p,
\end{equation}
respectively. The conservation equation in Eq.~\eqref{eq:conserv-general} with $\nabla_\nu T^{\mu\nu}=0$ features a single non-vanishing component. This equation can be further simplified using the chain rule $\dot{f}=f_R\dot{R}+f_T\dot{T}$, eliminating $\dot{R}$ and $\dot{T}$ using the first-order time derivatives of Eqs.~\eqref{eq:R} and \eqref{eq:T}, and using the result in Eq.~\eqref{eq:conserv-m} to eliminate $\dot{a}$. This results in the following simplified conservation equation:
\begin{equation}\label{eq:conserv-fG}
	f_T\left(\dot{p}-\dot{\rho}\right)=2\left(p+\rho\right)\dot{f_T}.
\end{equation}
At this point, it is important to note that the two field equations in Eqs.~\eqref{eq:Gtt} and \eqref{eq:Grr} along with the two conservation equations in Eqs.~\eqref{eq:conserv-m} and \eqref{eq:conserv-fG} are not linearly independent. Indeed, Eq.~\eqref{eq:Grr} can be obtained from a time derivative of Eq.~\eqref{eq:Gtt} followed by appropriade algebraic manipulations involving the remaining equations. One can thus discard one of these equations from the system. Due to its more complicated structure in comparison to the remaining equations, we chose to discard Eq. \eqref{eq:Grr}.

As outlined previously, to verify if sudden singularities are allowed in this formalism at some finite time $t_s$, we require $a$, $\dot{a}$ and $\rho$ to remain finite as $t\rightarrow t_s$. To preserve the regularity of $f\left(R,T\right)$, we also require this function, along with its partial derivatives and the variables $R$ and $T$, to remain finite as $t\rightarrow t_s$. From Eq.~\eqref{eq:R}, one verifies that the regularity of $R$ and $\dot a$ immediately forbids a divergence in $\ddot{a}$ (since there cannot be a divergence in one single term in an equation), whereas from Eq.~\eqref{eq:T} one verifies that the regularity of $T$ and $\rho$ forces the regularity of $p$. In turn, Eq. \eqref{eq:conserv-m} will then require $\dot{\rho}$ to be finite. In summary, $p$, $\dot{\rho}$, $\ddot{a}$, and all lower order time derivatives of the same quantities have to remain finite, and there are no sudden singularities at this order.

Let us now analyze the modified Friedmann equation in Eq.~\eqref{eq:Gtt}. To do so, it is necessary to expand the time derivatives of $f_R$ in terms of derivatives of $R$ and $T$. We note that time derivatives of the function $f(R,T)$ or of its partial derivatives $f_X$, where $X$ collectively denotes any combination of $R$ and $T$, can be expanded using the chain rule. That is, for a general partial derivative $f_X$:
\begin{equation}\label{eq:chain1}
	\dot{f_X}=f_{XR}\dot{R}+f_{XT}\dot{T},
\end{equation}
\begin{equation}\label{eq:chain2}
	\ddot{f_X}=f_{XR}\ddot{R}+f_{XT}\ddot{T}+f_{XRR}\dot{R}^2+f_{XTT}\dot{T}^2+2f_{XRT}\dot{R}\dot{T},
\end{equation}
and so on. Afterwards, the time derivatives of $R$ can be expressed in terms of derivatives of $a$ using Eq.~\eqref{eq:R}, whereas the derivatives of $T$ can be expressed as derivatives of $\rho$ and $p$ using  Eq.~\eqref{eq:T}. Taking Eq.~\eqref{eq:Gtt}, expanding $\dot{f}_R$ using the chain rule, and discarding the terms that are already required to remain finite and which would thus be subdominant in the limit $t\to t_s$, this asymptotic behaviour of Eq.~\eqref{eq:Gtt} can be written as
\begin{equation}\label{eq:aGtt}
2f_{RR}\frac{\dddot{a}}{a}+f_{RT}\dot{p}\simeq 0,
\end{equation}
i.e., since $f_{RR}$ and $f_{RT}$ remain finite, a divergence in $\dot{p}$ would have to be compensated by a divergence in $\dddot{a}$. However, following the same procedure, the asymptotic behaviour of Eq.~\eqref{eq:conserv-fG} becomes
\begin{equation}\label{eq:aconserv-fG}
\dot{p}\simeq \frac{12f_{RT}\left(p+\rho\right)}{f_T-6f_{TT}\left(p+\rho\right)}\left(\frac{\dddot{a}}{a}\right),
\end{equation}
which can be used to eliminate $\dot{p}$ in Eq.~\eqref{eq:aGtt}, thus leaving $\dddot{a}$ as the only variable left that is still allowed to diverge in this equation. In the absence of another quantity to counterbalance this divergence, one is forced to conclude that $\dddot{a}$ has to remain finite. Consequently, $\dot{p}$ is necessarily finite as well. This analysis proves that, assuming the regularity of the function $f\left(R,T\right)$ implies that no sudden singularities appear in the zero-order time derivatives of the system of cosmological equations, even though sudden singularities in higher-order derivatives of these equations are not yet excluded.

The most straightforward way to generalize the results obtained above to an arbitrary $n$th-order time derivative of the system of cosmological equations is to recur to the method of mathematical induction. For that purpose, we assume that the time derivatives $\rho^{(n)}$, $p^{(n)}$, and $a^{(n+2)}$ are finite, a feature already proven for $n=1$, and we show inductively that if the result holds for some order $n$ then it consequently holds for the following order $n+1$. In the notation used, the power in curved brackets symbolises the order of the time derivative. Taking then the $n$th-order time derivative of Eqs.~\eqref{eq:conserv-m}, \eqref{eq:Gtt} and \eqref{eq:conserv-fG}, using the chain rule to write time derivatives of $f_R$ and $f_T$ in terms of derivatives of $R$ and $T$, which are posteriorly written in terms of derivatives of $a$, $\rho$ and $p$ via Eqs.~\eqref{eq:R} and \eqref{eq:T}, taking the asymptotic limit $t\to t_s$ and discarding the non-dominant terms, one obtains the following system of asymptotic equations
\begin{equation}\label{eq:adnconserv-m}
	\rho^{\left(n+1\right)}\simeq0,
\end{equation}	
\begin{equation}\label{eq:adnGtt}
2f_{RR}\frac{a^{\left(n+3\right)}}{a}+f_{RT}p^{\left(n+1\right)}\simeq 0,
\end{equation}
\begin{equation}\label{eq:adnconserv-fG}
p^{\left(n+1\right)}\simeq \frac{12f_{RT}\left(p+\rho\right)}{f_T-6f_{TT}\left(p+\rho\right)}\left(\frac{a^{\left(n+3\right)}}{a}\right),
\end{equation}
respectively. One thus verifies that the $n$th-order time derivative of the matter conservation equation guarantees the regularity of $\rho^{(n+1)}$, whereas of the modified Friedmann equation and the second conservation equation in turn force the regularity of $a^{(n+3)}$ and $p^{(n+1)}$ by the same procedure as before. This argument can be extrapolated for any higher order derivative by a redefinition $N=n+1$, thus proving that, under the assumptions considered, sudden singularities are not allowed in any time derivative of $a$, $\rho$ and $p$.

%%%%%%%%%%%%%%%%%%%%%%%%%%%%%%%%%%%%%%%%%%%%%%%%%%%%%%%%%%%%%%%%%%%%%%%%
\subsection{Scalar-tensor representation}\label{subsec:typeII-scalar-tensor}
%%%%%%%%%%%%%%%%%%%%%%%%%%%%%%%%%%%%%%%%%%%%%%%%%%%%%%%%%%%%%%%%%%%%%%%%

Let us now turn to the scalar-tensor representation of the theory and perform a similar analysis. Under the assumptions detailed above, in Sec.~\ref{subsec:framework}, one obtains the following two independent field equations from Eq.~\eqref{eq:fields}, the modified Friedmann equation and the modified Raychaudhuri equation, which take the forms
\begin{equation}\label{eq:tt}
    \dot{\varphi}\left(\frac{\dot{a}}{a}\right) + \varphi \left( \frac{\dot{a}^{2} + k}{a^{2}} \right)  = \frac{8 \pi}{3}\rho + \frac{\psi}{6}\left( 3\rho - p\right)+ \frac{1}{6} V,
\end{equation}
\begin{equation}\label{eq:rr}
\begin{split}
   \ddot{\varphi}+ 2\dot{\varphi}\left(\frac{\dot{a}}{a}\right) +\varphi\left( \frac{2\ddot{a}}{a} + \frac{\dot{a}^{2}+k}{a^{2}} \right)  = -8\pi p  \\+ \frac{\psi}{2}\left(\rho-3p\right) + \frac{1}{2} V.
\end{split}
\end{equation}
 Furthermore, the equations of motion for the scalar fields $\varphi$ and $\psi$ from Eqs.~\eqref{eq:Vphi} and \eqref{eq:Vpsi} become
\begin{equation}\label{eq:Vphicosmo}
    V_{\varphi} = 6\left( \frac{\ddot{a}}{a}+ \frac{\dot{a}^{2}+k}{a^{2}}\right),
\end{equation}
\begin{equation}\label{eq:Vpsicosmo}
    V_{\psi} = 3p-\rho,
\end{equation}
respectively.
 Finally, the conservation equation from Eq.~\eqref{eq:conserv-general2} in this framework takes the form
\begin{equation}\label{eq:conserv-total}
\begin{multlined}
    8\pi(\rho+p)\left(\frac{\dot{a}}{a}\right)+\frac{8\pi}{3}\dot{\rho} = \dot{\varphi}\left(\frac{\ddot{a}}{a}+\frac{\dot{a}^2+k}{a^2}-\frac{1}{6}V_\varphi\right) \\
    -\dot{\psi}\left(\frac{1}{2}\rho - \frac{1}{6}p+\frac{1}{6}V_\psi\right) -\psi\left[\frac{\dot{a}}{a}(\rho+p)+\frac{1}{2}\dot{\rho} - \frac{1}{6}\dot{p}\right].
\end{multlined}
\end{equation}

The system of Eqs.~\eqref{eq:tt} to \eqref{eq:conserv-total} forms a system of five equations from which only four are linearly independent. To prove this feature, one can take the time derivative of Eq.~\eqref{eq:tt}, and use Eqs.~\eqref{eq:Vphicosmo} and \eqref{eq:Vpsicosmo} to eliminate the partial derivatives $V_\varphi$ and $V_\psi$. One then uses the conservation equation in Eq.~\eqref{eq:conserv-total} to eliminate the time derivative $\dot\rho$, and use the Raychaudhuri equation in Eq.~\eqref{eq:rr} to eliminate the second time derivative $\ddot a$, thus recovering the original equation. Thus, one of these equations can be discarded from the system without loss of generality. 

Inserting the result of Eq.~\eqref{eq:conserv-m} into the conservation equation in Eq.~\eqref{eq:conserv-total} and using Eqs.~\eqref{eq:Vphicosmo} and \eqref{eq:Vpsicosmo} to cancel the factors $V_\varphi$ and $V_\psi$, one obtains a simplified form of the general conservation equation as
\begin{equation}\label{eq:conserv-psi}
      \psi(\dot{p}-\dot{\rho})= 2(p+\rho)\dot{\psi},
\end{equation} 
which is analogous to Eq.~\eqref{eq:conserv-fG} in the geometrical representation. 

Therefore, we have a new system of six equations, namely, Eqs.~\eqref{eq:conserv-m}, \eqref{eq:tt}--\eqref{eq:Vpsicosmo},  and \eqref{eq:conserv-psi}, of which only five are linearly independent. Since one of the equations can be discarded without loss of generality, in this case we chose to discard Eq. \eqref{eq:conserv-psi} and work with the remaining. 

Similarly as before, to verify if sudden singularities can arise at some instant $t_s$ we assume $a$, $\dot{a}$, and $\rho$ to be finite. Furthermore, following the definition of the scalar fields $\varphi$ and $\psi$ as derivatives of $f\left(R,T\right)$ and recalling that we are assuming this function to be $C^\infty$, we also require $\varphi$ and $\psi$ to remain finite. Finally, as the potential $V\left(\varphi,\psi\right)$ along with its partial derivatives are constructed solely in terms of $\varphi$ and $\psi$, these must also be regular throughout the entire time evolution. Under these considerations, Eq.~\eqref{eq:Vpsicosmo} immediately forbids a divergence in $p$, and consequently Eq.~\eqref{eq:conserv-m} prevents a divergence in $\dot\rho$. Similarly, Eq. \eqref{eq:Vphicosmo} imposes the regularity of $\ddot a$. Thus, $p$, $\psi$, $\dot{\rho}$, $\ddot{\varphi}$, $\ddot{a}$ and all lower order derivatives are required to remain finite. Taking now into consideration the modified field equations, in the modified Friedmann equation in Eq.~\eqref{eq:tt} the only term still allowed to diverge is $\dot\varphi$. In the absence of a quantity to counterbalance this divergence, we are forced to conclude that $\dot\varphi$ must remain finite. Finally, the same argument holds for $\ddot\varphi$ in Eq.~\eqref{eq:rr}, which forces this quantity to remain finite. We have thus proven that no sudden singularities can arise in $\ddot a$, $\dot \rho$, $p$, $\ddot\varphi$, and $\psi$, or any lower order time derivative of the same quantities, even though divergences in higher-order time derivatives are still to be excluded.

Let us now proceed similarly as in the previous section and extend this analysis for any $n$th-order time derivative of these quantities via the method of mathematical induction. We thus assume that the time derivatives $\rho^{(n+1)}$, $p^{(n)}$, $a^{(n+2)}$, $\varphi^{(n+2)}$, and $\psi^{(n)}$ are regular, a feature already proven for $n=0$, and we show inductively that if the result holds for some order $n$ then it will remain true for the following order $n+1$ and, consequently, for any $n$. Taking the $n$th-order time derivative of Eqs.~\eqref{eq:conserv-m} and \eqref{eq:tt}--\eqref{eq:Vpsicosmo}, taking the asymptotic limit $t\to t_s$ and discarding the subdominant terms, i.e., all the terms assumed to be regular above, one is left with the following system of asymptotic equations:
  \begin{equation}\label{eq:adn+1conserv-m}
 \rho^{\left(n+2\right)}\simeq-3\frac{\dot{a}}{a}p^{\left(n+1\right)}.
 \end{equation}
 \begin{equation}\label{eq:adn+1tt}
 \psi p^{\left(n+1\right)}-\left(3\rho-p+V_{\psi}\right)\psi^{\left(n+1\right)}\simeq 0,
 \end{equation}
 \begin{equation}\label{eq:adn+1rr}
 \begin{multlined}
  2\varphi^{\left(n+3\right)}+4\varphi\frac{a^{\left(n+3\right)}}{a}+\left(16\pi+3\psi\right)p^{\left(n+1\right)}\\
  \simeq\left(\rho-3p+V_\psi\right)\psi^{\left(n+1\right)},
 \end{multlined}
 \end{equation}
 \begin{equation}\label{eq:adn+1Vphi}
 V_{\psi\varphi}\psi^{\left(n+1\right)}\simeq 6\frac{a^{\left(n+3\right)}}{a},
 \end{equation}
 \begin{equation}\label{eq:adn+1Vpsi}
 V_{\psi\psi}\psi^{\left(n+1\right)}\simeq 3 p^{\left(n+1\right)}.
 \end{equation}
One can thus verify that Eq.~\eqref{eq:adn+1Vpsi} can be used to eliminate $\psi^{\left(n+1\right)}$ from Eq.~\eqref{eq:adn+1tt}, where $p^{\left(n+1\right)}$ can now be factored out, yielding $p^{\left(n+1\right)}\simeq 0$. Hence, $p^{\left(n+1\right)}$ must remain finite, which implies by Eq.~\eqref{eq:adn+1Vpsi} that $\psi^{\left(n+1\right)}$ must also remain finite. Consequently, Eq. \eqref{eq:adn+1Vphi} prevents a divergence in $a^{\left(n+3\right)}$. Therefore, in Eq.~\eqref{eq:adn+1rr} the only remaining term allowed to diverge is $\varphi^{\left(n+3\right)}$, which in the absence of a quantity to counterbalance this divergence is forced to be regular. The same argument holds for $\rho^{\left(n+2\right)}$ in Eq.~\eqref{eq:adn+1conserv-m}. Again, this argument can be extrapolated to any higher order derivative via a redefinition ${N=n+1}$ and repeating the procedure, thus proving that in the framework considered no sudden singularities can arise in any time derivative of $a$, $\rho$, $p$, $\varphi$, and $\psi$ in the scalar-tensor representation of $f(R,T)$ gravity.

%%%%%%%%%%%%%%%%%%%%%%%%%%%%%%%%%%%%%%%%%%%%%%%%%%%%%%%%%%%%%%%%%%%%%%%%
\subsection{Implications to Type IV singularities}\label{subsec:typeIV}
%%%%%%%%%%%%%%%%%%%%%%%%%%%%%%%%%%%%%%%%%%%%%%%%%%%%%%%%%%%%%%%%%%%%%%%%

The previous subsections render several considerations regarding type IV singularities. If we include in this type of singularity the cases when $\rho$ and $p$ are simply finite (i.e., without requiring that ${\rho\rightarrow 0}$  and ${p\rightarrow 0}$ as ${t\rightarrow t_s}$), it was already seen that singularities do not appear in higher order derivatives of the scale factor. Still, the case in which ${\rho\rightarrow 0}$  and ${p\rightarrow 0}$ as ${t\rightarrow t_s}$ can be seen explicitly. Let us take the geometrical representation. If $a$, ${H=\dot{a}/a}$, and $R$ are required to be finite, then, by Eq.~\eqref{eq:R}, $\ddot{a}$ has to remain finite. Furthermore, if ${\rho\rightarrow 0}$  and ${p\rightarrow 0}$ as ${t\rightarrow t_s}$, then, by Eq.~\eqref{eq:conserv-m}, ${\dot{\rho}\rightarrow 0}$. Thus, Eq.~\eqref{eq:conserv-fG} now requires ${\dot{p}\rightarrow 0}$. As we have seen in previous subsections, if $\dot{\rho}$, $\dot{p}$, $\ddot{a}$, and all lower order time derivatives are all finite, no divergences are allowed to appear in higher orders time derivatives of the same quantities. Therefore, these type IV singularities are also prevented.

%%%%%%%%%%%%%%%%%%%%%%%%%%%%%%%%%%%%%%%%%%%%%%%%%%%%%%%%%%%%%%%%%%%%%%%%
\section{Sudden singularities with $\nabla_\nu T^{\mu\nu}\neq0$}\label{sec:suddens}
%%%%%%%%%%%%%%%%%%%%%%%%%%%%%%%%%%%%%%%%%%%%%%%%%%%%%%%%%%%%%%%%%%%%%%%%

In the previous analysis, one of the assumptions considered was the conservation of the stress-energy tensor, i.e., ${\nabla_\nu T^{\mu\nu}=0}$. This property is a mandatory feature in general relativity arising from the fact that the Einstein tensor $G_{\mu\nu}$ is divergence free, which results from the Bianchi identities; note that this feature also arises from the diffeomorphism invariance of the Einstein-Hilbert action \cite{Harko:2018ayt}. However, one may have that ${\nabla_\nu T^{\mu\nu} \neq 0}$ in specific modified theories of gravity, especially those with geometry-matter couplings \cite{Bertolami:2007gv,Koivisto:2005yk}. Indeed, in modified gravity it is usual to reorganise the gravitational field equations into an effective Einstein field equation  \cite{Capozziello:2014bqa,Capozziello:2013vna}, i.e., $G_{\mu\nu}=8\pi T_{\mu\nu}^{\text{eff}}$, where one defines a divergence-free effective stress-energy tensor $T_{\mu\nu}^{\text{eff}}$ containing not only the matter stress-energy tensor $T_{\mu\nu}$ but also contributions from the extra degrees of freedom of the gravitational sector, while allowing ${\nabla_\nu T^{\mu\nu}\neq0}$ \cite{Koivisto:2005yk}. In this section, we will thus explore the consequences of discarding the assumption of the conservation of the matter sector in the problem of sudden singularities.

We again assume that the universe is well-described by the FLRW metric given in Eq.~\eqref{eq:FLRW-metric}, and that matter is given by an isotropic perfect fluid, i.e., the stress-energy tensor $T_{\mu\nu}$ and the auxiliary tensor $\Theta_{\mu\nu}$ are given by Eqs.~\eqref{eq:em-fluid} and \eqref{eq:Theta-fluid}, respectively. Furthermore, we still assume that the function $f\left(R,T\right)$ is an arbitrary $C^\infty$ function, which consequently implies that both $R$ and $T$ remain finite throughout their entire time evolution.

%%%%%%%%%%%%%%%%%%%%%%%%%%%%%%%%%%%%%%%%%%%%%%%%%%%%%%%%%%%%%%%%%%%%%%%%
\subsection{Sudden singularities in $\dddot a$: geometrical representation}
\label{subsec:sudden-ddda-geom}
%%%%%%%%%%%%%%%%%%%%%%%%%%%%%%%%%%%%%%%%%%%%%%%%%%%%%%%%%%%%%%%%%%%%%%%%

The system of equations that describes this framework is given by the field equations in Eqs.~\eqref{eq:Gtt} and \eqref{eq:Grr}, alongside with the general conservation equation given in Eq.~\eqref{eq:conserv-general}. Since the general conservation equation is not linearly independent of the two field equations, we choose to discard this equation and work solely with the two field equations. Similarly as in the situation studied in Sec. \ref{subsec:typeII-geometrical}, the regularity of $R$ and $\dot a$ in Eq.~\eqref{eq:R} forces $\ddot a$ to be finite, whereas the regularity of $\rho$ and $T$ in Eq.~\eqref{eq:T} forces $p$ to remain finite. Thus, in the limit $t\to t_s$, the asymptotic forms of Eqs.~\eqref{eq:Gtt} and \eqref{eq:Grr} become
\begin{equation}\label{eq:asymp1}
\dot f_R\simeq0,
\end{equation}
\begin{equation}\label{eq:asymp2}
2\dot f_R\frac{\dot a}{a}+\ddot f_R\simeq0.
\end{equation}
The time derivatives of $f_R$ in Eqs.~\eqref{eq:asymp1} and \eqref{eq:asymp2} can be extended in terms of derivatives of $R$ and $T$ via the chain rules in Eqs.~\eqref{eq:chain1} and \eqref{eq:chain2}, which in turn can be converted into derivatives of $a$, $\rho$, and $p$ via Eqs.~\eqref{eq:R} and \eqref{eq:T}. From Eq.~\eqref{eq:asymp1}, one will obtain terms proportional solely to $\dddot a$, $\dot \rho$, and $\dot p$, whereas Eq.~\eqref{eq:asymp2} will also feature terms proportional to $a^{(4)}$, $\ddot \rho$, and $\ddot p$. Due to the large number of possibly divergent quantities, one or more equations are needed to close the system. Since we have chosen to work solely with the field equations, we shall take the derivative of Eq.~\eqref{eq:Gtt} as an extra equation, which takes the asymptotic form
\begin{equation}
\frac{1}{2}\dot f -3f_R\frac{\dddot a}{a} +3\ddot f_R \frac{\dot a}{a} \simeq 8\pi \dot \rho+\dot f_T\left(\rho+p\right)+f_T\left(\dot \rho+\dot p\right).\label{eq:asymp3}
\end{equation}

We are now in conditions to prove the possibility of sudden singularities in this framework. Equation \eqref{eq:asymp1} can be solved for $\dot p$ and inserted into Eq.~\eqref{eq:asymp2}, which in turn can be solved for $\ddot p$ and both replaced into Eq.~\eqref{eq:asymp3}. Given the precise dependence of Eqs.~\eqref{eq:asymp2} and \eqref{eq:asymp3} in $a^{(4)}$ and $\ddot \rho$, these two quantities are eliminated from the system automatically upon the manipulation described. One is thus left with an equation relating $\dot \rho$ with $\dddot a$ of the form
\begin{eqnarray}\label{eq:allowssudden}
\dot \rho\left(8\pi+\frac{4}{3}f_T\right)\simeq\frac{\dddot a}{a}\Bigg\{&&\frac{f_{RR}}{f_{RT}}\left[6\left(\rho+p\right)f_{TT}-f_T\right]
	\nonumber \\
&&-6\left(\rho+p\right)f_{RT}\Bigg\}.
\end{eqnarray}
Without any extra equation to provide a different relationship between $\dot \rho$ and $\dddot a$, one concludes that sudden singularities may arise in this theory at the level of the third time-derivative of the scale factor, sourced by a diverging $\dot\rho$. If one now imposes the conservation of matter, i.e., Eq.~\eqref{eq:conserv-m}, $\dot\rho$ is forced to be finite, which consequently imposes that $\dddot a$ remains finite and one recovers the case studied in Sec.~\ref{subsec:typeII-geometrical}.

Note however that Eq.~\eqref{eq:allowssudden} is only valid if $f_{RT}\neq 0$, as the first step towards its derivation requires a division by this factor. Thus, the particular case $f_{RT}=0$ must be considered independently. If $f_{RT}=0$, Eq.~\eqref{eq:asymp1} will force $\dddot a$ to remain finite, whereas Eq.~\eqref{eq:asymp2} in turn forces $a^{(4)}$ to remain finite. From Eq.~\eqref{eq:asymp3}, one verifies that a divergence in $\dot \rho$ and $\dot p$ is still possible as these two quantities can compensate each other, but this divergence does not incur with any effects on the scale factor, thus lacking relevance in a cosmological context.

%%%%%%%%%%%%%%%%%%%%%%%%%%%%%%%%%%%%%%%%%%%%%%%%%%%%%%%%%%%%%%%%%%%%%%%%
\subsection{Sudden singularities in $\dddot a$: scalar-tensor representation} \label{subsec:sudden-ddda-ST}
%%%%%%%%%%%%%%%%%%%%%%%%%%%%%%%%%%%%%%%%%%%%%%%%%%%%%%%%%%%%%%%%%%%%%%%%

Complementary, in the scalar-tensor representation, the system of equations that describes this framework is given by the field equations in Eqs.~\eqref{eq:tt} and \eqref{eq:rr}, alongside with the general conservation equation given in Eq.~\eqref{eq:conserv-total}. Again, since the general conservation equation is not linearly independent of the two field equations, we choose to discard this equation and work solely with the two field equations. Additionally, there is also the two extra equations from the scalar fields in Eqs.~\eqref{eq:Vphicosmo} and \eqref{eq:Vpsicosmo}. Similar to the situation studied in Sec.~\ref{subsec:typeII-scalar-tensor}, the regularity of $V_\varphi$ and $\dot a$ in Eq.~\eqref{eq:Vphicosmo} forces $\ddot a$ to be finite, whereas the regularity of $\rho$ and $V_\psi$ in Eq.~\eqref{eq:Vpsicosmo} forces $p$ to remain finite. Since we also have that $\varphi$, $\psi$ and $V$ are regular, in the limit $t\to t_s$, the asymptotic forms of Eqs.~\eqref{eq:tt} and \eqref{eq:rr} become
\begin{equation}\label{eq:asymp-ST-nonconserv-tt}
\dot{\varphi} \simeq 0,
\end{equation}
\begin{equation}\label{eq:asymp-ST-nonconserv-rr}
\ddot{\varphi}+ 2\dot{\varphi}\left(\frac{\dot{a}}{a}\right) \simeq 0.
\end{equation}
Thus, Eq.~\eqref{eq:asymp-ST-nonconserv-tt} immediately requires $\dot{\varphi}$ to be finite. Hence, Eq~\eqref{eq:asymp-ST-nonconserv-rr} forces $\ddot{\varphi}$ to be regular. So, all remains finite at this order, but it can be shown that singularities can appear in the following order, i.e., with $\dddot{a}$. Taking the time derivative of Eqs.~\eqref{eq:tt}--\eqref{eq:Vpsicosmo} one obtains the following asymptotic equations, in the limit $t\to t_s$,
\begin{equation}\label{eq:asymp-ST-nonconserv-dtt}
	  \frac{8 \pi}{3}\dot{\rho} + \frac{\dot{\psi}}{6}\left( 3\rho - p\right)+ \frac{\psi}{6}\left( 3\dot{\rho} - \dot{p}\right)+ \frac{1}{6} V_\psi\dot{\psi} \simeq 0,
\end{equation}
\begin{equation}\label{eq:asymp-ST-nonconserv-drr}
	 \dddot{\varphi} +2\varphi\frac{\dddot{a}}{a}  \simeq -8\pi \dot{p} + \frac{\dot{\psi}}{2}\left(\rho-3p\right)+ \frac{\psi}{2}\left(\dot{\rho}-3\dot{p}\right) + \frac{1}{2} V_\psi \dot{\psi},
\end{equation}
\begin{equation}\label{eq:asymp-ST-nonconserv-dVphi}
	V_{\varphi\psi} \dot{\psi} \simeq 6 \frac{\dddot{a}}{a},
\end{equation}
\begin{equation}\label{eq:asymp-ST-nonconserv-dVpsi}
	V_{\psi\psi}\dot{\psi} \simeq 3\dot{p}-\dot{\rho},
\end{equation}
respectively. Using Eq.~\eqref{eq:asymp-ST-nonconserv-dVphi} to eliminate $\dot{\psi}$ and Eq.~\eqref{eq:asymp-ST-nonconserv-dVpsi} to eliminate $\dot{p}$, Eqs.~\eqref{eq:asymp-ST-nonconserv-dtt} and \eqref{eq:asymp-ST-nonconserv-drr} become
\begin{equation}\label{eq:asymp-ST-nonconserv-dtt2}
	\dot{\rho}\left(8\pi+\frac{4}{3}\psi\right) \simeq \frac{\dddot{a}}{a}\left(\frac{3\left(p-3\right)\rho +\psi V_{\psi\psi} -3V_\psi}{V_{\varphi\psi}}\right),
\end{equation}
\begin{eqnarray}\label{eq:asymp-ST-nonconserv-drr2}
	\dddot{\varphi} &\simeq & \frac{\dddot{a}}{a}\left[\frac{3\left(\rho-3p\right)-\left(16\pi+3\psi\right)V_{\psi\psi}+3V_\psi}{V_{\varphi\psi}}-2\varphi\right]
	\nonumber \\
	&&  \qquad -\frac{8\pi}{3}\dot{\rho},
\end{eqnarray}
respectively. Eliminating $\dot{\rho}$ from the last equation, simply provides a relationship between $\dddot{\varphi}$ and $\dddot{a}$,
\begin{eqnarray}\label{eq:asymp-ST-nonconserv-drr3}
&&\dddot\varphi \simeq \frac{1}{V_{\varphi\psi}}\frac{\dddot a}{a}\bigg\{-\frac{8\pi\left[3\left(p-3\rho\right)+\psi V_{\psi\psi}-3V_\psi\right] }{4\left(6\pi+\psi\right)} 
%\qquad 
\nonumber\\
 && \; + 3\left(\rho-3p\right)-\left(16\pi+3\psi\right)V_{\psi\psi}+3V_\psi-2\varphi V_{\varphi\psi}\Big\}.
\end{eqnarray}

Both $\dddot a$ and $\dddot\varphi$ are still allowed to diverge. Therefore, without any extra equation, one concludes that sudden singularities may arise at the level of the third time-derivative of the scale factor, as seen in the geometrical representation case. At the same time, if one now imposes the conservation of matter, i.e., Eq.~\eqref{eq:conserv-m}, $\dot\rho$ is forced to be finite, which consequently imposes that $\dddot a$ remains finite [by Eq.~\eqref{eq:asymp-ST-nonconserv-dtt2}]. Then $\dot\psi$, $\dot p$ and $\dddot\varphi$ have also to be regular [by Eqs.~\eqref{eq:asymp-ST-nonconserv-dVphi}, \eqref{eq:asymp-ST-nonconserv-dVpsi} and \eqref{eq:asymp-ST-nonconserv-drr3}, respectively] and one recovers the case studied in Sec.~\ref{subsec:typeII-scalar-tensor}.

%%%%%%%%%%%%%%%%%%%%%%%%%%%%%%%%%%%%%%%%%%%%%%%%%%%%%%%%%%%%%%%%%%%%%%%%
\subsection{Model featuring a sudden singularity}\label{subsec:modelsingularity}
%%%%%%%%%%%%%%%%%%%%%%%%%%%%%%%%%%%%%%%%%%%%%%%%%%%%%%%%%%%%%%%%%%%%%%%%

In the previous subsections, we have proven that if one considers a situation in which the matter stress-energy tensor is not conserved, sudden singularities might appear in the third-order time derivative of the scale factor. In this section, we will thus provide an explicit model for which sudden singularities arise. In the framework considered, the system of cosmological equations (in the geometrical representation) is composed by two independent field equations, namely Eqs.~\eqref{eq:Gtt} and \eqref{eq:Grr}, for the three unknowns $a$, $\rho$ and $p$. This implies that one can still impose an extra constraint to close the system. 

We thus choose to impose a particular form of the scale factor \cite{Barrow:2004xh} given by
\begin{equation}\label{eq:scale}
a\left(t\right)=\left(a_s-1\right)\left(\frac{t}{t_s}\right)^\gamma+1-\left(1-\frac{t}{t_s}\right)^\delta,
\end{equation}
where the parameter $a_s$ represents the scale factor at the instant $t_s$ when the sudden singularity occurs, and $\gamma$ and $\delta$ are constant exponents that must satisfy a few constraints in order to provide a sudden singularity at the right order of the time derivatives of the scale factor. Indeed, the $n$th-order time derivative of Eq.~\eqref{eq:scale} is given by
\begin{eqnarray}
&&a^{(n)}\left(t\right)=\left(a_s-1\right)\left(\frac{t}{t_s}\right)^{\gamma-n}t_s^{-n}\prod_{i=0}^{n-1}\left(\gamma-i\right)
	\label{dscale1} 
	\nonumber \\
&&\qquad \quad +\left(-1\right)^{n+1}\left(1-\frac{t}{t_s}\right)^{\delta-n}t_s^{-n}\prod_{i=0}^{n-1}\left(\delta-i\right).
\end{eqnarray}

Taking the limit $t\to 0$, one verifies that the first term on the right-hand side of Eq.~\eqref{dscale1} guarantees the regularity of all time derivatives up to order $n$ at the initial time $t=0$, provided that $\gamma>n$. On the other hand, the second term on the right-hand side of Eq.~\eqref{dscale1} is responsible for causing the sudden singularity to appear at $t=t_s$ precisely at order $n$ while maintaining all derivatives of lower order regular, provided that $n-1<\delta<n$. Furthermore, both parameters $\delta$ and $\gamma$ are required to be non-whole, to avoid the problematic situations $\delta=n$ and $\gamma=n$, which would effectively cancel the necessary time dependencies of $a^{\left(n\right)}$. Since we are interested in sudden singularities appearing at $n=3$, we consider the parameters $\delta$ and $\gamma$ within the following bounds: $2<\delta<3<\gamma<4$ (we consider $\gamma<4$ to prevent $\gamma$ from being an integer within the considered bounds).

\begin{figure}
	\includegraphics[scale=0.7]{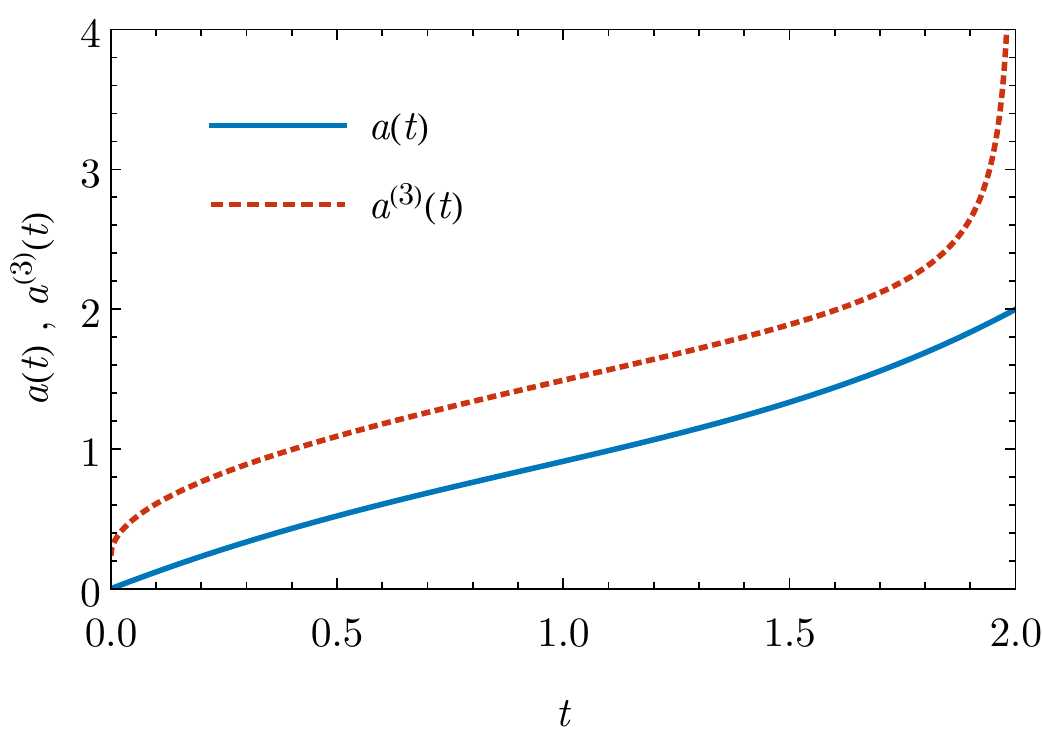}
	\caption{Scale factor $a\left(t\right)$ from Eq.~\eqref{eq:scale} and third-order time derivative $\dddot a$ from Eq.~\eqref{dscale1} with $n=3$, $a_s=t_s=2$, $\delta=2.5$, and $\gamma=3.5$. The scale factor remains finite throughout the entire time evolution, whereas $\dddot a$ diverges as $t\to t_s$.}
	\label{fig:scale}
\end{figure}

In Fig.~\ref{fig:scale} we plot the scale factor $a\left(t\right)$ and its third-order time derivative $\dddot a\left(t\right)$ for $a_s=t_s=2$, $\delta=2.5$ and $\gamma=3.5$. The scale factor has an initial deceleration period followed by a late-time cosmic acceleration. From this figure, it is clear that the scale factor remains finite throughout the entire time evolution, whereas $\dddot a$ diverges at the singularity time $t_s$, as necessary.

%%%%%%%%%%%%%%%%%%%%%%%%%%%%%%%%%%%%%%%%%%%%%%%%%%%%%%%%%%%%%%%%%%%%%%%%
\subsection{Constraints from the cosmological parameters}\label{subsec:cosmo-constraints}
%%%%%%%%%%%%%%%%%%%%%%%%%%%%%%%%%%%%%%%%%%%%%%%%%%%%%%%%%%%%%%%%%%%%%%%%

The scale factor and the time at the singularity, $a_s$ and $t_s$, respectively, can be constrained via the comparison of our results with the experimental measurements of the cosmological parameters, in particular the present-time values of the Hubble parameter $H_0$ and the deceleration parameter $q_0$, respectively. These two functions are defined in terms of the scale factor and its time derivatives as
\begin{equation}\label{cosmopar}
H=\frac{\dot a}{a}, \qquad q=-\frac{\ddot a a}{\dot a^2}.
\end{equation}

\begin{figure*}
\includegraphics[scale=0.55]{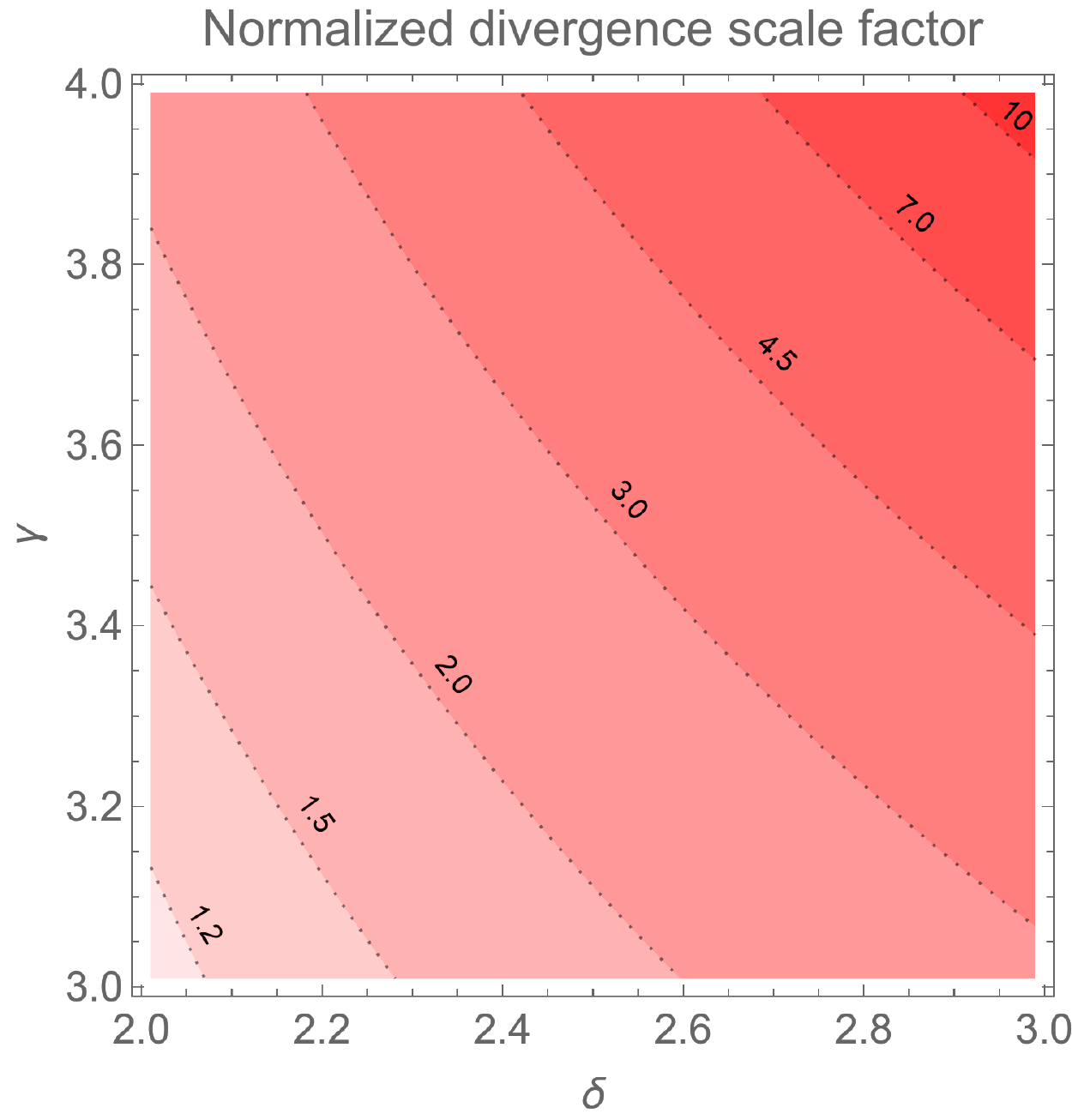}\qquad \qquad
\includegraphics[scale=0.55]{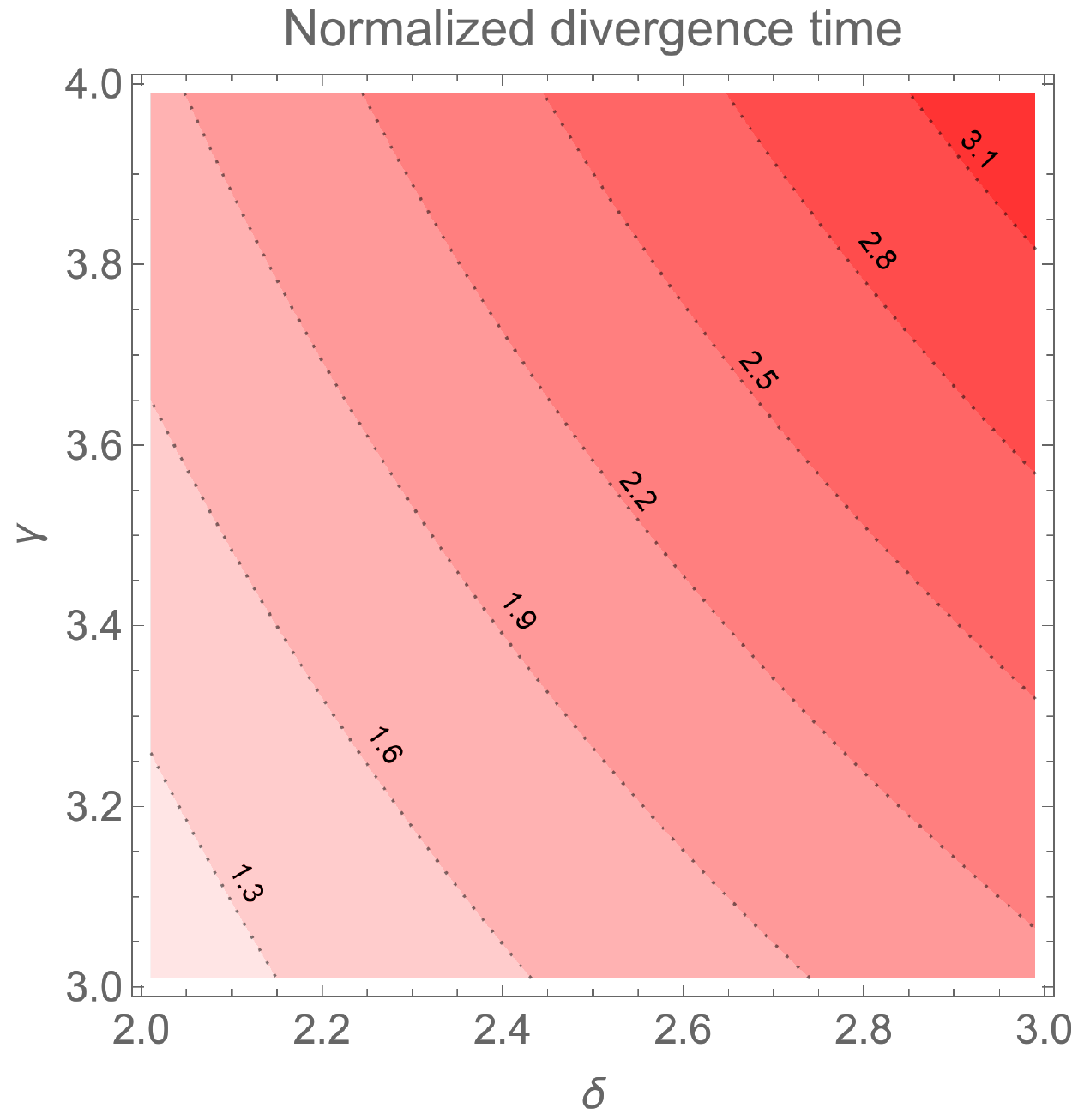}
\caption{Normalized divergence scale factor $\bar a=a_s/a_0$ (left panel) and Normalized divergence time $\bar t=t_s/t_0$ (right panel) as a function of $\delta$ and $\gamma$ for $H=H_0$, $q=q_0$ and $t=t_0$. Both $\bar a$ and $\bar t$ are shown to increase with $\delta$ and $\gamma$.}
\label{fig:divergence}
\end{figure*}

According to the most recent Cosmic Microwave Background (CMB) measurements from the \textit{Planck} satellite \cite{Planck:2018vyg}, the experimental values of these parameters at the present time (assuming a flat-$\Lambda$CDM model) are ${H_0 = 67.4 	\pm 0.5 \text{ km s}^{-1}\text{Mpc}^{-1}}$ and ${q_0=-0.527 \pm 0.011}$ (using ${q_0=\frac{1}{2}\Omega_{m0}-\Omega_{\Lambda0}}$, where $\Omega_{m0}$ and $\Omega_{\Lambda0}$ are the present-time density parameters of matter and dark energy in the form of a cosmological constant, respectively). The age of the universe, as also measured by \textit{Planck} 2018 (TT, TE, EE+lowE+lensing), is ${t_0 = 13.797 \pm 0.023 \text{ Gy}}$. Inserting the scale factor from Eq.~\eqref{eq:scale} into Eq.~\eqref{cosmopar}, one obtains 
\begin{equation}\label{hubble}
H\left(t\right)=\frac{\gamma\left(a_s-1\right)\left(t-t_s\right)\left(\frac{t}{t_s}\right)^\gamma-\delta t\left(1-\frac{t}{t_s}\right)^\delta}{t\left(t-t_s\right)\left[\left(a_s-1\right)\left(\frac{t}{t_s}\right)^\gamma+1-\left(1-\frac{t}{t_s}\right)^\delta\right]},
\end{equation}
\begin{eqnarray}\label{deceleration}
q\left(t\right)&=&-\frac{\left(a_s-1\right)\left(\frac{t}{t_s}\right)^\gamma+1-\left(1-\frac{t}{t_s}\right)^\delta}{\left[\gamma\left(a_s-1\right)\left(t-t_s\right)\left(\frac{t}{t_s}\right)^\gamma-\delta t\left(1-\frac{t}{t_s}\right)^\delta\right]^2} 
	\nonumber \\
&&\qquad \times \Bigg[\gamma\left(\gamma-1\right)\left(a_s-1\right)\left(t-t_s\right)^2\left(\frac{t}{t_s}\right)^\gamma
	\nonumber  \\
&&\qquad \qquad \qquad  -\delta\left(\delta-1\right)t^2\left(1-\frac{t}{t_s}\right)^\delta\Bigg].
\end{eqnarray}
Equations \eqref{hubble} and \eqref{deceleration} can be used to impose constraints on the values of $a_s$ and $t_s$ that are consistent with the cosmological observations. To do so, one takes the limit $t\to t_0$ in these equations and introduces the experimental values of $H_0$, $q_0$ and $t_0$. The result is a system of two equations for the fours unknowns $a_s$, $t_s$, $\delta$ and $\gamma$. Since the parameters $\delta$ and $\gamma$ are constrained by the inequalities $2<\delta<3<\gamma<4$, one can now specify particular values of $\delta$ and $\gamma$, i.e., consider different models for the scale factor, and compute the associated solutions for $a_s$ and $t_s$. In Fig.~\ref{fig:divergence} we plot the normalized divergence scale factor ${\bar a \equiv a_s/a_0}$ and the normalized divergence time ${\bar t \equiv t_s/t_0}$, where we have defined $a_0=a\left(t=t_0\right)$, as a function of $\delta$ and $\gamma$, where the constraints $H=H_0$ and $q=q_0$ and $t=t_0$ were taken into consideration. Both parameters $\bar a$ and $\bar t$ are shown to increase with $\delta$ and $\gamma$, with the values of $a_s$ and $t_s$ ranging from minimum values of $a_s\sim 1.0651 a_0$ and $t_s\sim 1.06736 t_0$ in the limit $\delta\to 2$ and $\gamma\to 3$, to maximum values of $a_s\sim 11.6976 a_0$ and $t_s\sim 3.33272 t_0$ in the limit  $\delta\to 3$ and $\gamma\to 4$. 

\begin{figure*}
\includegraphics[scale=0.55]{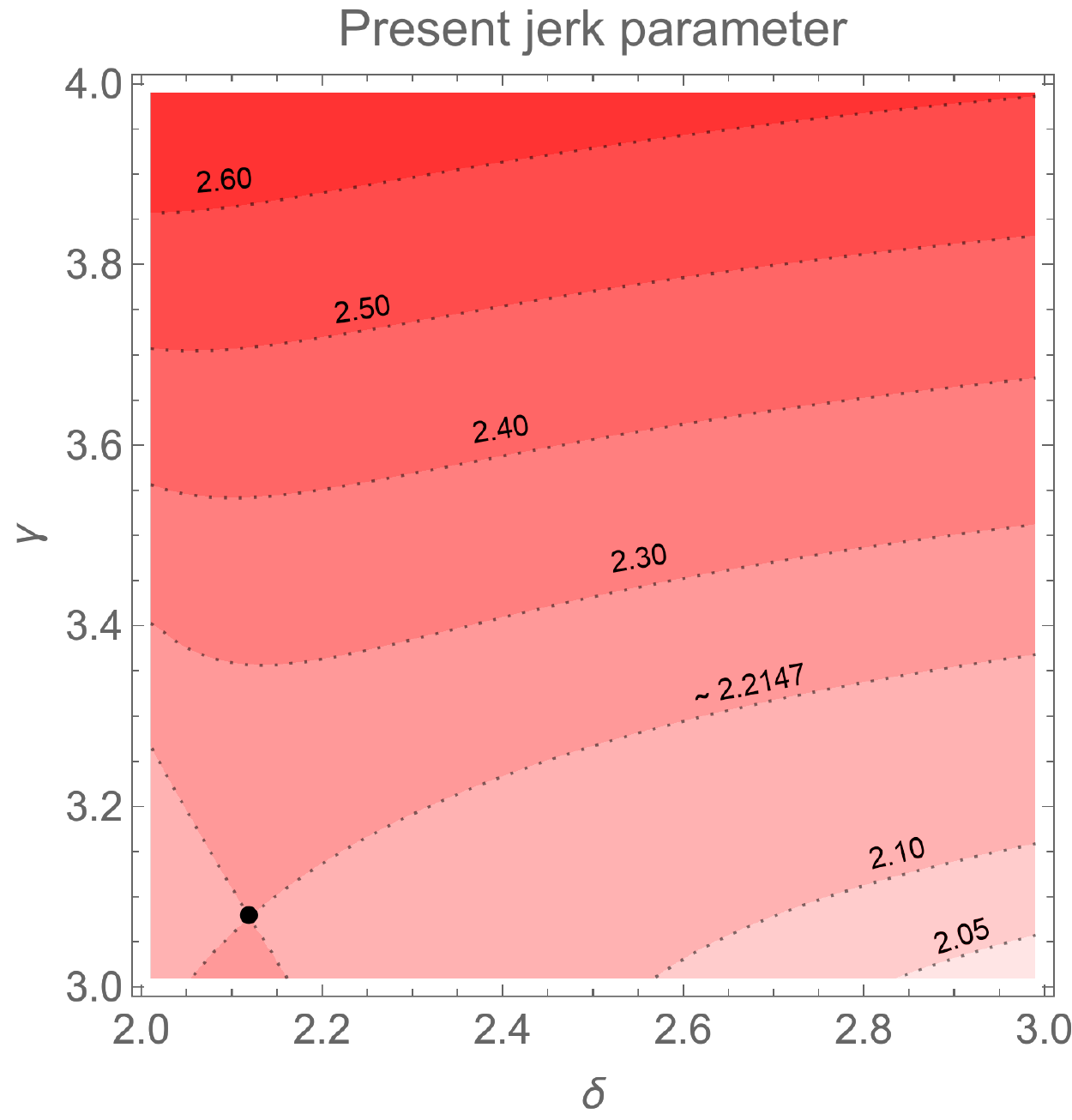}\qquad \qquad
\includegraphics[scale=0.55]{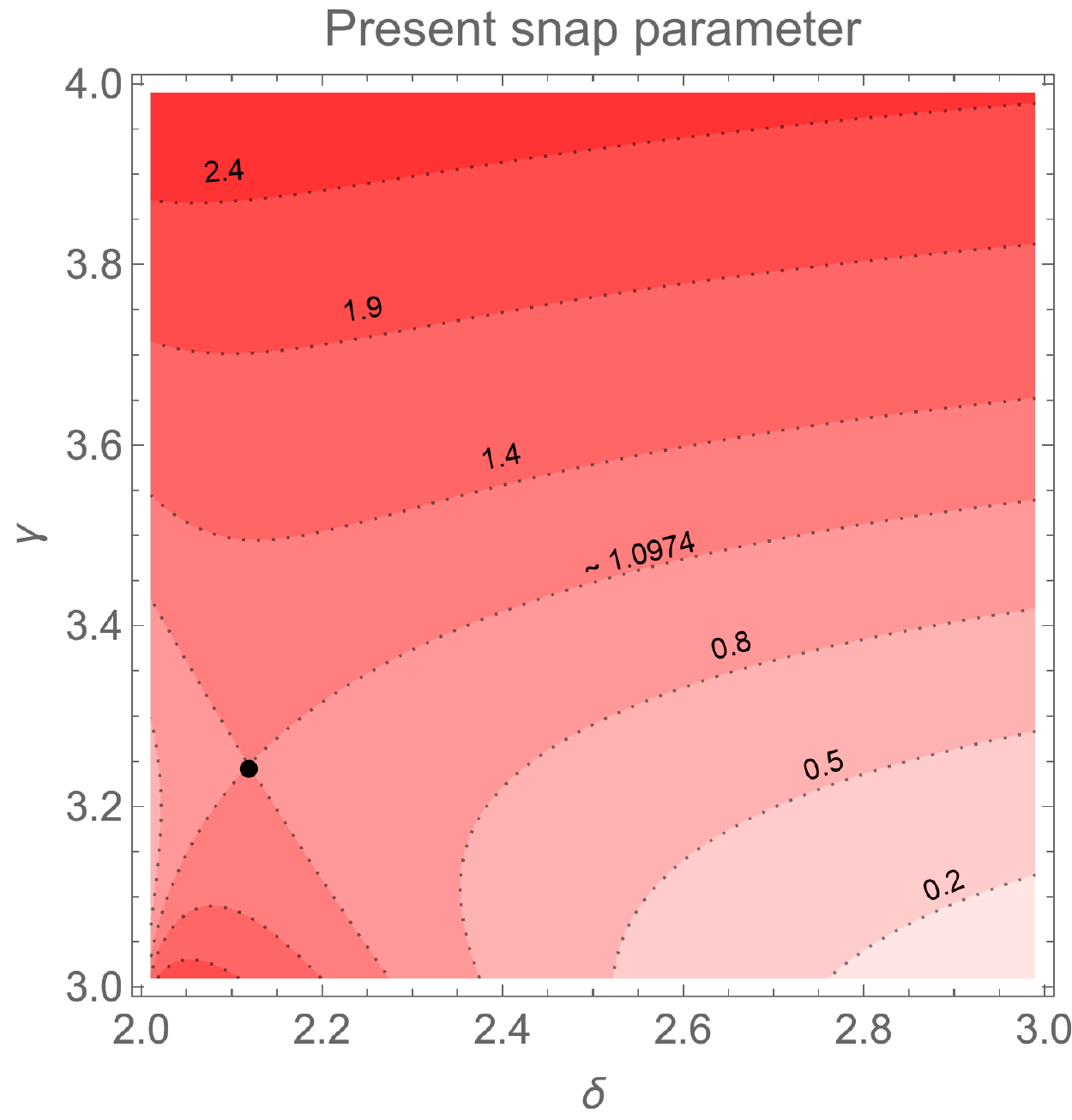}
\caption{Predictions for the present jerk parameter $j_0$ (left panel) and the present snap parameter $s_0$ (right panel) as a function of $\delta$ and $\gamma$ for $H=H_0$, $q=q_0$ and $t=t_0$ and with the values of $a_s$ and $t_s$ consistent with the ones from Fig.~\ref{fig:divergence}.}
\label{fig:prediction}
\end{figure*}

Finally, one can also use this framework to provide predictions for the cosmological jerk and snap parameters, which we denote by $j$ and $s$, respectively. These parameters are constructed in terms of higher-order time derivatives of the scale factor as \cite{Lobo:2020hcz}
\begin{equation}\label{cosmopar2}
j=\frac{\dddot a a^2}{\dot a^3},\qquad s=\frac{a^{(4)}a^3}{\dot a^4}.
\end{equation}
Inserting the scale factor from Eq.~\eqref{eq:scale} into Eqs.~\eqref{cosmopar2}, specifying the values of the exponents $\delta$ and $\gamma$ as well as the corresponding values of $a_s$ and $t_s$ arising from the analysis shown in Fig.~\ref{fig:divergence}, and taking the limit ${t\to t_0}$, one obtains a prediction for the current values of these cosmological parameters, i.e., $j_0=j\left(t=t_0\right)$ and $s_0=s\left(t=t_0\right)$. In Fig.~\ref{fig:prediction}, we plot the predictions for the present jerk parameter $j_0$ and present snap parameter $s_0$ as a function of $\delta$ and $\gamma$ and consistent with the analysis described previously. In the range of parameters allowed, the present jerk parameter $j_0$ and the present snap parameter $s_0$ range from minimum values of $j_0\sim 2.02229$ and $s_0\sim 0$ in the limit $\delta\to 3$ and $\gamma\to 3$ and maximum values of $j_0\sim 2.69639$ and $s_0\sim 2.84078$ in the limit $\delta\to 2$ and $\gamma\to 4$. 


%%%%%%%%%%%%%%%%%%%%%%%%%%%%%%%%%%%%%%%%%%%%%%%%%%%%%%%%%%%%%%%%%%%%%%%%
\subsection{Energy conditions and constraints on $f(R,T)$}\label{subsec:E-conditions}
%%%%%%%%%%%%%%%%%%%%%%%%%%%%%%%%%%%%%%%%%%%%%%%%%%%%%%%%%%%%%%%%%%%%%%%%

The energy conditions are a set of inequalities for the matter fields $\rho$ and $p$ that guarantee the physical relevance of the solutions considered. In summary, for a perfect fluid considered in this work, the null energy condition (NEC) requires that ${\rho+p \geq 0}$, the weak energy condition (WEC) implies that ${\rho+p \geq 0}$ and ${\rho \geq 0}$, the strong energy condition (SEC) entails that ${\rho+3p \geq 0}$ and ${\rho \geq 0}$, and the dominant energy condition (DEC) imposes that ${\rho \geq |p|}$. The models featuring sudden singularities introduced in the previous subsections are characterized by finite $\rho$ and $p$, but no constraints on their values are required. On the other hand, these models are also characterized by a divergent $\dot\rho$, but not a necessarily divergent $\dot p$. Thus, a positive divergence of $\dot\rho$ works in favor of the energy conditions, whereas a negative divergence works against it. The analysis of the requirements on the free parameters of the model that satisfy the energy conditions at the singularity time $t_s$, i.e., that imply a positive divergence in $\dot\rho$, allows one to impose constraints on specific models of $f\left(R,T\right)$ gravity. For concreteness, let us consider a simple form of the $f\left(R,T\right)$ gravity consisting of a simple extension of GR with a crossed term on $R$ and $T$, given by
\begin{equation}\label{particualrformfRT}
f\left(R,T\right)=R+\alpha RT,
\end{equation}
for some constant parameter $\alpha$. Note that, according to the results of Sec. \ref{subsec:sudden-ddda-geom}, the crossed term is essential to guarantee the existence of a sudden singularity in $\dddot{a}$. For this particular choice of $f\left(R,T\right)$, Eq.~\eqref{eq:allowssudden} becomes 
\begin{equation}\label{eq:allowssudden-particular-f}
\dot \rho\left(8\pi+\frac{4}{3}\alpha R\right)\simeq -6\alpha \frac{\dddot a}{a}\left(\rho+p\right).
\end{equation}

For the scale factor given in Eq.~\eqref{eq:scale}, Fig.~\ref{fig:scale} shows that $a(t_s)>0$ and ${\dddot{a}(t_s) \to + \infty >0}$. Furthermore, for the NEC to be satisfied, the factor $\left(\rho+p\right)$ is also positive. Thus, to guarantee that $\dot\rho>0$, one needs the following constraint to be satisfied: 
\begin{equation}\label{eq:alpha-condition}
\frac{1}{\alpha}\left(8\pi+\frac{4}{3}\alpha R\right)<0.
\end{equation}
For the FLRW spacetime, the Ricci scalar takes the form ${R=6(\ddot{a}a+\dot{a}^2+k)/a^2}$ which, with the scale factor given in Eq.~\eqref{eq:scale} and in the limit ${t \to t_s}$, becomes 
\begin{equation}\label{constraintR}
R(t_s) = \frac{6}{a_s^2t_s^2} \left\{\gamma \left(a_s-1\right)\left[\left(2\gamma - 1\right)a_s - \gamma\right] + kt_s^2\right\}.
\end{equation}

\begin{figure*}
	\includegraphics[scale=0.55]{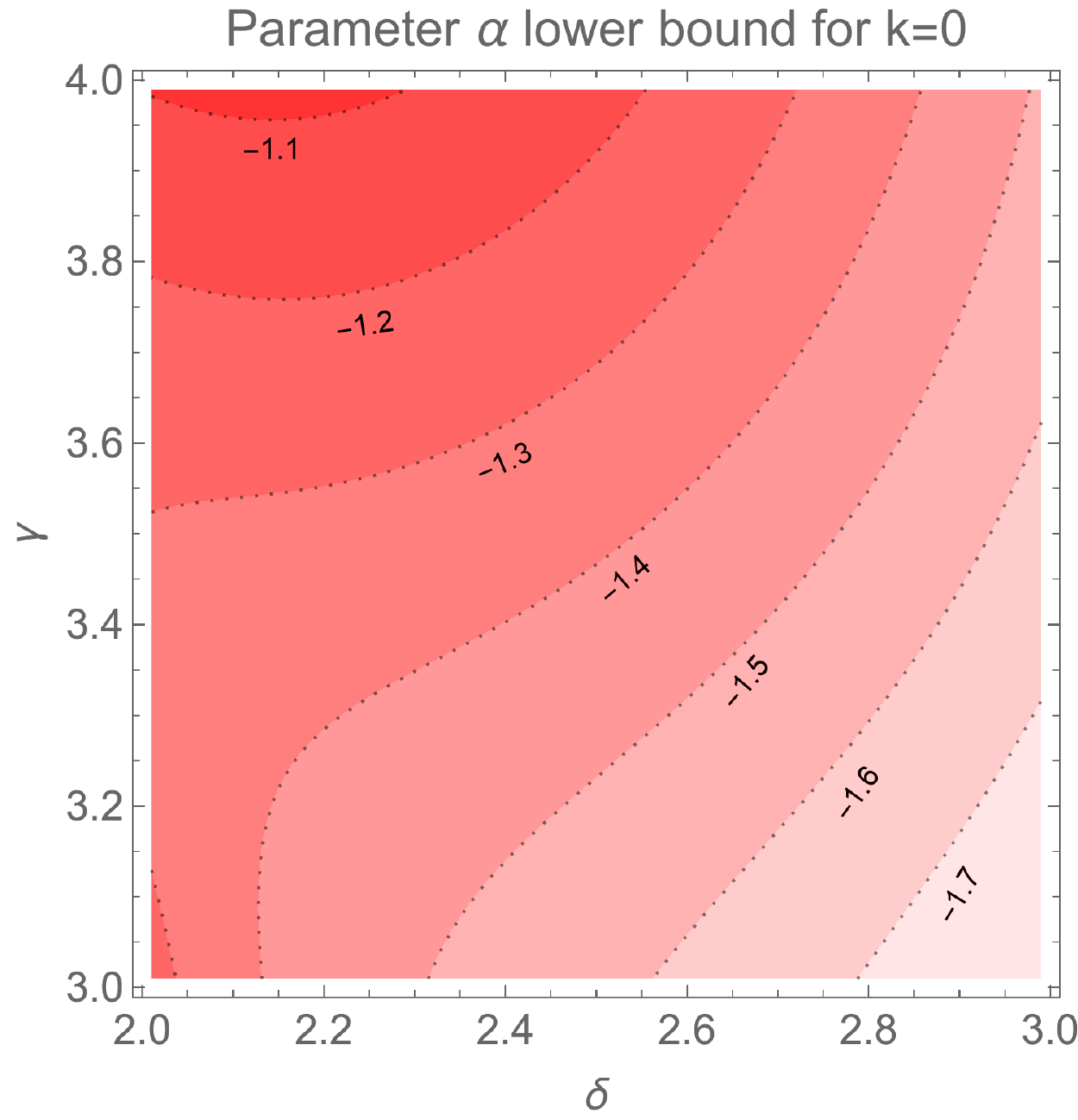} \\
	\includegraphics[scale=0.55]{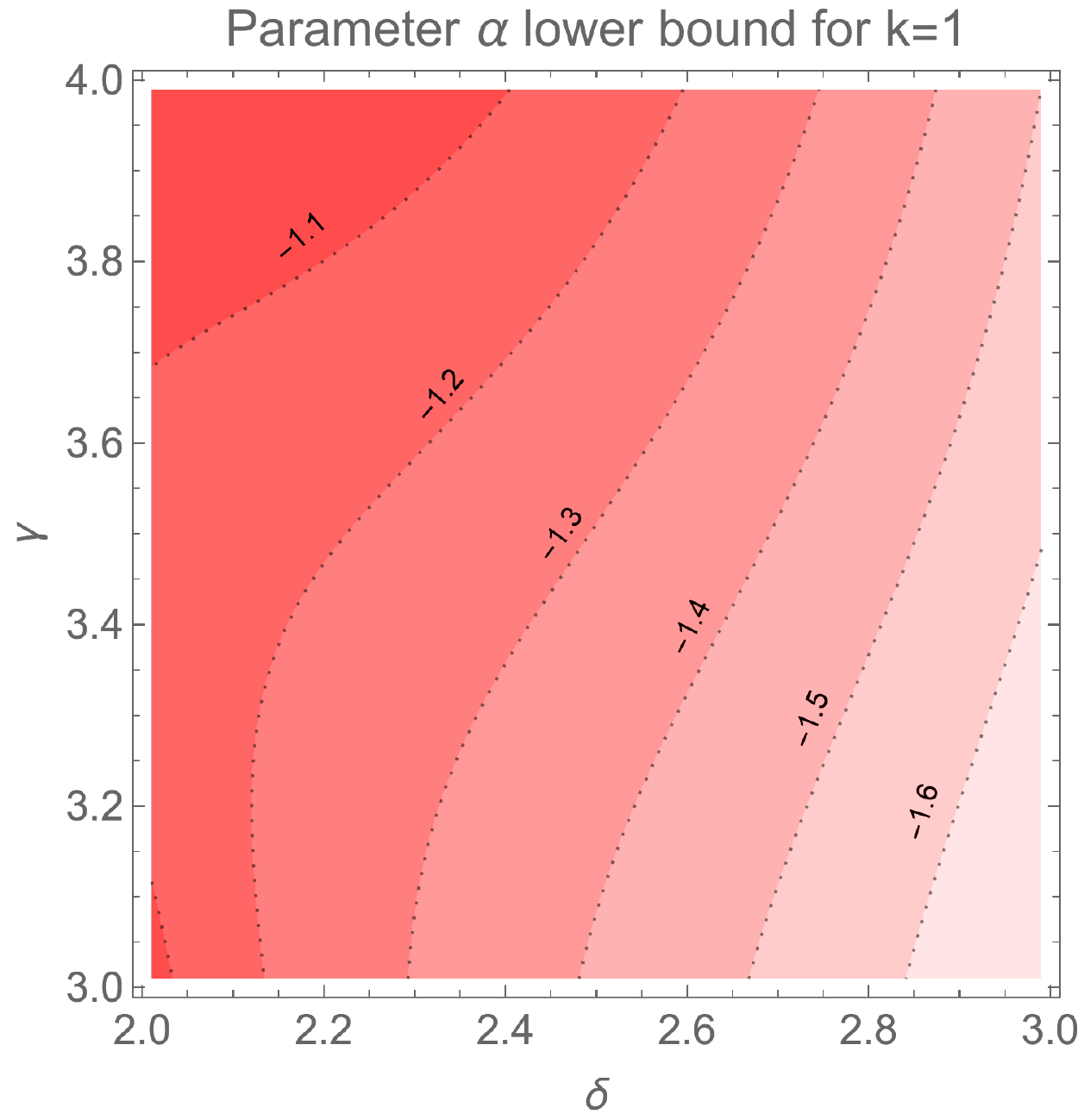}\qquad \qquad 
	\includegraphics[scale=0.55]{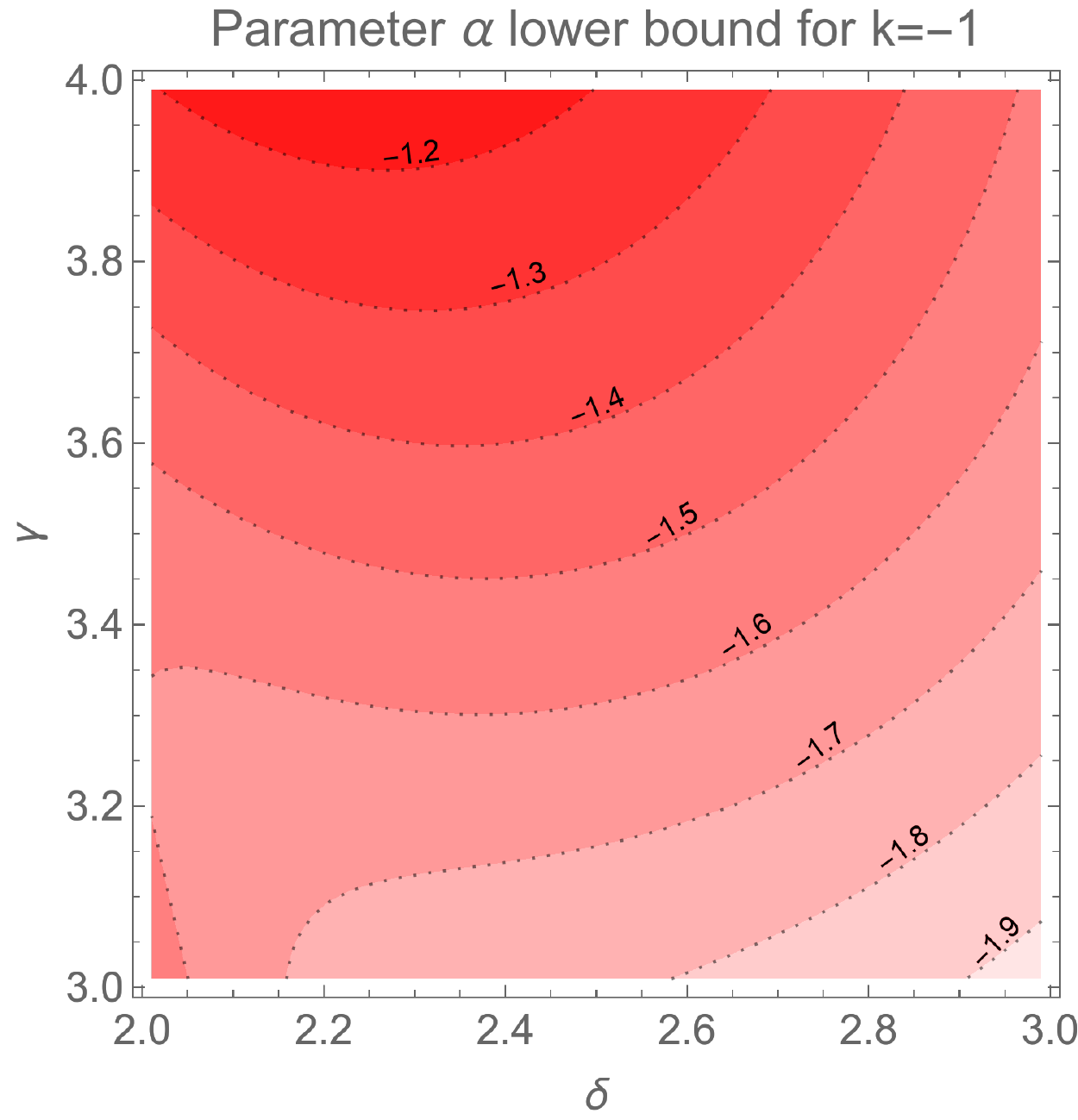}
	\caption{Lower bounds on the parameter $\alpha$ for the range $2<\delta<3<\gamma<4$ and for the three values of the curvature parameter ${k=\{-1,0,1\}}$, in the particular model with ${f(R,T)=R+\alpha RT}$ and the scale factor as given in Eq.~\eqref{eq:scale}, from energy conditions considerations. Note that the upper bound is always ${\alpha<0}$, according to Eq~\eqref{eq:alpha-conditions-partclr}. As with Fig.~\ref{fig:divergence}, these results were obtained with the experimental values of $H_0$, $q_0$ and $t_0$ as used in Sec.~\ref{subsec:cosmo-constraints}.}
	\label{fig:alpha}
\end{figure*}

In the previous subsection, we obtained the values of $a_s$ and $t_s$ that are consistent with the cosmological parameters for the ranges of $\delta$ and $\gamma$ considered, ${2<\delta<3<\gamma<4}$. One can thus think of $a_s$ and $t_s$ as unique functions of the parameters $\delta$ and $\gamma$, i.e., $a_s\left(\delta,\gamma\right)$ and $t_s\left(\delta,\gamma\right)$. Consequently, the Ricci scalar in Eq.~\eqref{constraintR} is completely determined by a combination of $\delta$ and $\gamma$, for each specific value of $k=\{-1,0,1\}$. Furthermore, one verifies that $R\left(r_s\right)$ is always positive. Consequently, from Eq.~\eqref{eq:alpha-condition} one obtains the constraint on the parameter $\alpha$
\begin{equation}\label{eq:alpha-conditions-partclr}
-\frac{6\pi}{R(t_s)} < \alpha < 0.
\end{equation}
  One can thus determine the numerical values of the lower bound on $\alpha$ in Eq.~\eqref{eq:alpha-conditions-partclr}, for the parameter space under consideration $2<\delta<3<\gamma<4$. The results are shown in Fig.~\ref{fig:alpha} for $k=\{-1,0,1\}$.

The function $f\left(R,T\right)$ given in Eq.~\eqref{particualrformfRT} satisfies the requirement ${f_{RR}f_{TT}\neq f_{RT}^2}$ as long as $\alpha\neq 0$ and thus it has a well-defined equivalent scalar-tensor representation. For completeness, we shall also present what is the form of the potential $V\left(\varphi,\psi\right)$ that corresponds to the scalar-tensor version of the function $f\left(R,T\right)$ in Eq.~\eqref{particualrformfRT}. Taking the partial derivatives of $f\left(R,T\right)$, one verifies that $\varphi$ and $\psi$ are given by $\varphi=\alpha T+1$ and $\psi=\alpha R$. These definitions allow us to write $R$ and $T$ as a function of $\varphi$ and $\psi$, respectively. Thus, the potential $V\left(\varphi,\psi\right)$ in Eq.~\eqref{eq:potential} takes the form
\begin{equation}
V\left(\varphi,\psi\right)=V_0\psi\left(\varphi-1\right),
\end{equation}
where $V_0=1/\alpha$. The same constraints found in Eq.~\eqref{eq:alpha-conditions-partclr} can be found in the scalar-tensor representation using a potential of the form described in the previous equation.

%%%%%%%%%%%%%%%%%%%%%%%%%%%%%%%%%%%%%%%%%%%%%%%%%%%%%%%%%%%%%%%%%%%%%%%%
\section{Conclusions}\label{sec:conclusion}
%%%%%%%%%%%%%%%%%%%%%%%%%%%%%%%%%%%%%%%%%%%%%%%%%%%%%%%%%%%%%%%%%%%%%%%%

In this work, we have studied the possibility of sudden singularities arising in FLRW universes populated by isotropic perfect-fluid matter in the framework of the $f\left(R,T\right)$ gravity theory in both the geometrical and the scalar-tensor representations. We have searched for sudden singularities occurring in a finite-time future instant $t_s$, where the pressure of the fluid $p$ is allowed to diverge, while the energy density $\rho$, the expansion scale factor $a$ and the Hubble function ${H=\dot{a}/a}$ all remain finite throughout the entire time evolution. If one assumes the conservation of the stress-energy tensor of the matter sector, i.e., $\nabla_\nu T^{\mu\nu}=0$, in both representations of the theory we have proven that if $f(R,T)$ is taken to be a general C$^\infty$ function, and hence regular throughout the entire time evolution, no sudden singularities of the type described above can arise, forcing not only $p$ but also $\ddot a$, $\dddot a$, $\dot \rho$, and $\dot p$ to remain regular.

Furthermore, we used the methods of mathematical induction to extend the validity of the conclusions traced in the previous paragraph to any arbitrary $n$th-order time derivative of the quantities $p$, $\rho$ and $a$, thus proving that no sudden singularities are allowed in either of the representations of the theory for any higher-order time derivatives. This result contrasts with what was found in other modified gravity theories with two extra scalar degrees of freedom, such as in the generalized hybrid metric-Palatini gravity \cite{Rosa:2021ish}, in which divergences in higher-order time derivatives might arise even if they are prevented in their lower-order counterparts. 

Due to the similarities between sudden singularities (type II) and singularities of type IV, our results can be straightforwardly extrapolated to singularities of the latter type. Analyzing the possibility of type IV singularities appearing at some finite time instant $t_s$ due to divergences in higher derivatives of $a$ while $a\rightarrow a_s$, $\rho\rightarrow 0$ and $\left|p\right|\rightarrow 0$, and even including the cases in which $\rho$ and $p$ approach some asymptotically non-zero value, one verifies that these singularities are not allowed under the framework in study, according to our results. 

The conclusions traced above were facilitated by two main assumptions in the framework: the function $f\left(R,T\right)$ must remain regular throughout the entire time evolution, and the stress-energy tensor of the perfect fluid must be conserved, i.e., $\nabla_\nu T^{\mu\nu}=0$. We note, however, that none of these assumptions is mandatory. In particular, dropping the assumption that the matter sector must fulfill the conservation equation, the two independent conservation equations in Eqs.~\eqref{eq:conserv-m} and \eqref{eq:conserv-fG} merge into a single equation, thus increasing the number of degrees of freedom and allowing for sudden singularities to arise at the level of the third-order time derivative of the scale factor.

For the situations in which sudden singularities arise in $\dddot a$, we have provided an explicit example of a cosmological model for which $\dddot a$ diverges at some singularity instant $t_s$, while all other lower-order derivatives of $a$ remain finite throughout the entire time evolution. A comparison of our results with the experimental measurements of the Hubble constant, deceleration parameter, and age of the universe by the \textit{Planck} satellite allowed us to impose constraints on the values of the divergence time and divergence scale factor consistent with the experimental observations. Furthermore, our framework allowed us to provide predictions for the currently still unmeasured cosmological jerk and snap parameters, the first with a value of roughly $j_0\sim 2$, and the second ranging from $s_0\sim 0$ to $s_0\sim 3$. 

Finally, requiring that the system evolves in a direction that favors the validity of the energy conditions at the divergence time $t_s$, we were able to impose constraints on a particular model of $f\left(R,T\right)$ gravity that extends GR with a single crossed term of the form $\alpha RT$. In particular, we have proven that the coupling constant $\alpha$ must be negative, with a lower bound that is roughly between $-1$ and $-2$ depending on the values of the parameters $\delta$ and $\gamma$. Unlike the cosmological parameters, the lower bound on $\alpha$ is affected in a non-negligible manner by the curvature parameter $k$, the bounds being the strongest for $k=1$ and weakest for $k=-1$.

%%%%%%%%%%%%%%%%%%%%%%%%%%%%%%%%%%%%%%%%%%%%%%%%%%%%%%%%%%%%%%%%%%%%%%%%
\begin{acknowledgments}
%%%%%%%%%%%%%%%%%%%%%%%%%%%%%%%%%%%%%%%%%%%%%%%%%%%%%%%%%%%%%%%%%%%%%%%%
T.B.G. and F.S.N.L. acknowledge support from the Funda\c{c}\~{a}o para a Ci\^{e}ncia e a Tecnologia  (FCT) research grants UIDB/04434/2020 and UIDP/04434/2020, and through the FCT projects with reference PTDC/FIS-OUT/29048/2017 and PTDC/FIS-AST/0054/2021.
J.L.R. was supported by the European Regional Development Fund and the programme Mobilitas Pluss (MOBJD647).
F.S.N.L. also acknowledges support from the Funda\c{c}\~{a}o para a Ci\^{e}ncia e a Tecnologia (FCT) Scientific Employment Stimulus contract with reference CEECINST/00032/2018, and funding from the research grant CERN/FIS-PAR/0037/2019. 
\end{acknowledgments}
%%%%%%%%%%%%%%%%%%%%%%%%%%%%%%%%%%%%%%%%%%%%%%%%%%%%%%%%%%%%%%%%%%%%%%%%

%%%%%%%%%%%%%%%%%%%%%%%%%%%%%%%%%%%%%%%%%%%%%%%%%%%%%%%%%%%%%%%%%%%%%%%%

%%%%%%%%%%%%%%%%%%%%%%%%%%%%%%%%%%%%%%%%%%%%%%%%%%%%%%%%%%%%%%%%%%%%%%%%
\end{document}